\documentclass[aps,11pt]{revtex4}
\usepackage{amsmath,amssymb,mathptmx,slashed}
\usepackage[pdftex]{graphicx}
\newcommand{\Lx}{\left(}
\newcommand{\Rx}{\right)}
\newcommand{\LB}{\left[}
\newcommand{\RB}{\right]}

\newcommand{\ep}{{\varepsilon}}
\newcommand{\eqn}[1]{Eq.\,(\ref{#1})}
\newcommand{\eqns}[2]{Eqs.\,(\ref{#1}-\ref{#2})}
\newcommand{\fig}[1]{Fig.\,(\ref{#1})}

\newcommand{\app}[1]{Appendix~\ref{#1}}

\newcommand{\lo}{{LO}}
\newcommand{\nlo}{{NLO}}
\newcommand{\nnlo}{{NNLO}}
\newcommand{\nnnlo}{{N${}^3$LO}}
\newcommand{\qcd} {{QCD}}

\newcommand{\cdr} {{CDR}}
\newcommand{\hv} {{HV}}
\newcommand{\fdh} {{FDH}}
\newcommand{\dred} {{DRED}}
\newcommand{\ms} {{MS}}
\newcommand{\dr} {{DR}}
\newcommand{\msbar} {{\overline{\rm MS}}}
\newcommand{\drbar} {{\overline{\rm DR}}}
\newcommand{\asbare} {{\alpha_s^{B}}}
\newcommand{\aphibare} {{\alpha_\phi^{B}}}
\newcommand{\aebare} {{\alpha_e^{B}}}
\newcommand{\avbare} {{\alpha_{V}^{B}}}
\newcommand{\av} {{\alpha_{V}}}
\newcommand{\asbarepi} {{\Lx\frac{\alpha_s^{B}}{\pi}\Rx}}
\newcommand{\aebarepi} {{\Lx\frac{\alpha_e^{B}}{\pi}\Rx}}

\newcommand{\avebare} {{\alpha_{Ve}^{B}}}
\newcommand{\amsbar} {{\alpha_{s}^{\overline{\rm\ms}}}}
\newcommand{\amsbarpi} {{\Lx\frac{\alpha_{s}^{\overline{\rm\ms}}}{\pi}\Rx}}
\newcommand{\aphimsbar} {{\alpha_{\phi}^{\overline{\rm\ms}}}}

\newcommand{\aphidrbar} {{\alpha_{\phi}^{\overline{\rm\dr}}}}

\newcommand{\betabar}[1] {{\beta_{#1}^{\overline{\rm\ms}}}}
\newcommand{\betadred}[1] {{\beta_{#1}^{\overline{\rm\dr}}}}
\newcommand{\betaedred}[2] {{\beta_{e,\,#1\,#2}^{\overline{\rm\dr}}}}
\newcommand{\betavedred}[2] {{\beta_{Ve,\,#1\,#2}^{\overline{\rm\dr}}}}
\newcommand{\betafdh}[1] {{\beta_{#1}^{\overline{\rm\fdh}}}}
\newcommand{\gammabar}[1] {{\gamma_{#1}^{\overline{\rm\ms}}}}
\newcommand{\gammadred}[1] {{\gamma_{#1}^{\overline{\rm\dr}}}}

\newcommand{\adrbar} {{\alpha_{s}^{\overline{\rm\dr}}}}
\newcommand{\adrbarpi} {{\Lx\frac{\alpha_{s}^{\overline{\rm\dr}}}{\pi}\Rx}}
\newcommand{\aedrbar} {{\alpha_{e}^{\overline{\rm\dr}}}}
\newcommand{\etadrbar}[1] {{\eta_{#1}^{\drbar}}}
\newcommand{\etadrbarpi}[1] {{\Lx\frac{\eta_{#1}^{\drbar}}{\pi}\Rx}}
\newcommand{\aedrbarpi} {{\Lx\frac{\alpha_{e}^{\overline{\rm\dr}}}{\pi}\Rx}}
\newcommand{\avedrbar} {{\alpha_{Ve}^{\overline{\rm\dr}}}}

\newcommand{\afdhbar} {{\alpha_{s}^{\overline{\rm\fdh}}}}
\newcommand{\afdhbarpi} {{\Lx\frac{\alpha_{s}^{\overline{\rm\fdh}}}{\pi}\Rx}}
\newcommand{\Tr}[1]{\mathop{\rm Tr\LB{#1}\RB}\nolimits}
\newcommand{\mtrxelm}[3] {{{\left\langle#1\left|#2\right|#3\right\rangle}}}

\begin{document}
%\begin{titlepage}

\today

\bibliographystyle{apsrev}

\title{Regularization Schemes and Higher Order Corrections}

\author{William~B.~Kilgore}
\affiliation{Physics Department, Brookhaven National Laboratory,
  Upton, New York
  11973, USA.\\
  {\tt [kilgore@bnl.gov]} }

\begin{abstract}
I apply commonly used regularization schemes to a multiloop
calculation to examine the properties of the schemes at higher orders.
I find complete consistency between the conventional dimensional
regularization scheme and dimensional reduction, but I find that the
four-dimensional helicity scheme produces incorrect results at
next-to-next-to-leading order and singular results at
next-to-next-to-next-to-leading order.  It is not, therefore, a
unitary regularization scheme.
\end{abstract}

\maketitle

\section{Introduction}
Dimensional regularization~\cite{'tHooft:1972fi} is an elegant and
efficient means of handling the divergences that arise in perturbation
theory beyond the tree level.  Among its many favorable qualities it
respects gauge and Lorentz invariance and allows one to handle both
ultraviolet and infrared divergences in the same manner.  The
application of dimensional regularization to different kinds of
problems has led to the development of a variety of regularization
schemes, which share the dimensional regularization of momentum
integrals, but differ in their handling of external (or observed)
states and of spin degrees of freedom.

The original formulation of dimensional
regularization~\cite{'tHooft:1972fi}, known as the 't~Hooft-Veltman
(\hv) scheme, specifies that observed states are to be treated as
four-dimensional, while internal states are to be treated as $D_m = 4
- 2\,\ep$ dimensional.  That is, both their momenta and spin degrees
of freedom were to be continued from four to $D_m$ dimensions.  It
turns out that one has the freedom to choose the value of the trace of
the Dirac unit matrix to take its canonical value of four, so fermions
continue to have two spin degrees of freedom, even though their
momenta are continued to $D_m$ dimensions.  Internal gauge bosons,
however, have $D_m-2$ spin degrees of freedom (internal massive gauge
bosons have $D_m-1$ degrees of freedom).

A slight variation on the \hv\ scheme has come to be called
conventional dimensional regularization (\cdr)~\cite{Collins:Renorm}.
In this variation, all particles and momenta are taken to be $D_m$
dimensional.  This often turns out to be computationally more
convenient, since one set of rules governs all interactions.  This is
particularly so when computing higher order corrections to theories
subject to infrared sensitivities, like \qcd.  In the \hv\ scheme, if
two external states have infrared sensitive overlaps, they must be
treated as internal, or $D_m$ dimensional states.  In the \cdr\
scheme, all states are already treated as $D_m$ dimensional, so there
is no possibility of failing to properly account for infrared
overlaps.

A third variation, called dimensional reduction
(\dred)~\cite{Siegel:1979wq}, was devised for application to
supersymmetric theories.  In supersymmetry, it is essential that the
number of bosonic degrees of freedom is exactly equal to the number of
fermionic degrees of freedom.  This requirement is violated in the
\hv\ and \cdr\ schemes.  In the \dred\ scheme, the continuation to
$D_m$ dimensions is taken as a compactification from four dimensions.
Thus, while space-time is taken to be four-dimensional and particles
have the standard number of degrees of freedom, momenta span a $D_m$
dimensional vector space and momentum integrals are regularized
dimensionally.

A fourth variation, the four-dimensional helicity (\fdh)
scheme~\cite{Bern:1992aq,Bern:2002zk}, was developed primarily for use
in constructing one-loop amplitudes from unitarity cuts.  The most
efficient building blocks for such calculations are tree-level
helicity amplitudes, which necessarily have two spin degrees of
freedom for both fermions and gauge bosons.  The \fdh\ scheme
resembles the \dred\ scheme in that it regularizes momentum integrals
dimensionally while maintaining the spin degrees of freedom of a
four-dimensional theory (and therefore appears to be a valid
supersymmetric regularization scheme~\cite{Bern:2002zk}), but there
are crucial differences, which I will discuss in detail.

The fact that the \hv\ scheme respects the unitarity of the $S$-matrix
was proven at its introduction~\cite{'tHooft:1972fi}.  The arguments
which establish the validity of the \hv\ scheme carry over to the
\cdr\ scheme and establish that it too is a valid regularization
scheme. After some initial confusion over the proper renormalization
procedure~\cite{vanDamme:1984ig,Capper:1980ns,Jack:1993ws} for the
\dred\ scheme, it was established that it too is a proper, unitary
regularization scheme~\cite{Jack:1993ws} and that it is indeed
equivalent to the \cdr\ scheme~\cite{Jack:1994bn}.  The \fdh\ scheme
has never been subjected to such stringent examination.  It has been
used successfully in a number of landmark next-to-leading order (\nlo)
calculations, but it has never been established whether it is a
proper, unitary regularization scheme, or merely a set of shortcuts
that allow expert users to obtain correct results.

In this paper, I will perform a well-known multiloop calculation in
the various regularization schemes.  I will show that while the \hv\
and \cdr\ scheme calculations yield the correct result and the \dred\
scheme calculation, while far more complicated is completely
equivalent, the \fdh\ scheme calculation yields incorrect results
which inevitably violate unitarity at sufficiently high order.  A
detailed comparison of the various calculations identifies the source
of the unitarity violations in the \fdh\ scheme.

The plan of this paper is as follows: in section two, I will describe
the test calculation to be performed and present the result to be
obtained.  In sections three, four and five, I will describe in detail
the calculation to next-to-next-to-leading order (\nnlo) as it is
performed in the \cdr, \dred\ and \fdh\ schemes, respectively.  In
section six, I present partial results at \nnnlo\ which solidify the
conclusion that the \cdr\ and \dred\ schemes are equivalent and
correct, but that the \fdh\ scheme violates unitarity.  In section
seven, I will discuss my results and draw my conclusions.

\section{The Test Environment}
To test the regularization schemes, I will calculate two quantities:
the massless nonsinglet contributions to 
\begin{enumerate}
\item the hadronic decay width of a fictitious neutral vector boson
  $V$, of mass $M_V$; 
\item the single photon approximation to the total hadronic
  annihilation cross section for an electron -- positron pair. 
\end{enumerate}
I will perform these calculations by means of the optical theorem,
taking the imaginary part of the forward scattering amplitudes.  In
both cases, this means taking the imaginary part of the vacuum
polarization tensor sandwiched between external states.  Since the
optical theorem is a direct consequence of the unitarity of the
$S$-matrix, any unitary regularization scheme must give the same
result, once one expands in terms of a standard coupling.  To avoid
complications involving prescriptions for handling $\gamma_5$ and the
Levi-Civita tensor, I will take $V$ to have only vectorlike
couplings.  In this way, the vacuum polarization tensor for the $V$
boson will be identical to that of the off shell photon, up to
coupling constants and so the \qcd\ expansion of the two results will
differ only by constant numerical factors.

Each regularization scheme will start from the same four-dimensional
Lagrangian,
\begin{equation}
\begin{split}
{\cal L} =& - \frac{1}{2}A^{a}_{\mu}\Lx\partial^{\mu}\partial^{\nu}
   (1-\xi^{-1}) - g^{\mu\nu}\Box\Rx A^{a}_{\nu} -
g\,f^{abc}(\partial^\mu\,A^{a\,\nu})A^b_\mu\,A^c_\nu
- \frac{g^2}{4}f^{abc}\,f^{ade}\,A^{b\,\mu}\,A^{c\,\nu}\,A^d_\mu\,A^e_\nu\\
& + i\sum_f\,
\overline{\psi}_f^{i}\Lx\delta_{ij}\slashed{\partial}
    -i\,g\,t^{a}_{ij}\slashed{A}^a
    -i\,g_V\,Q_f\slashed{V}\Rx\,\psi_f^{j}
  - \overline{c}^{a}\Box\,c^a
    + g\,f^{abc}\Lx\partial_\mu\,\overline{c}^a\Rx\,A^{b\,\mu}\,c^c\,,
\label{eqn::4dlagrange}
\end{split}
\end{equation}
where $A^{a\,\mu}$ is the \qcd\ gauge field, $V^\mu$ is the massive
vector boson, $\psi_f$ is the quark field of flavor $f$,
$\overline{c}^a$ and $c^a$ are the Faddeev-Popov ghost fields, $g$ is
the \qcd\ coupling, $g_V$ is the $V$ gauge coupling and $Q_f$
represents the charge of the quark flavor $f$ under the $V$ symmetry.
I will not be computing nontrivial corrections in $g_V$, so there is
no need to specify the $V$-self interaction parts of the Lagrangian.

\begin{figure}[h]
\includegraphics[width=4.cm]{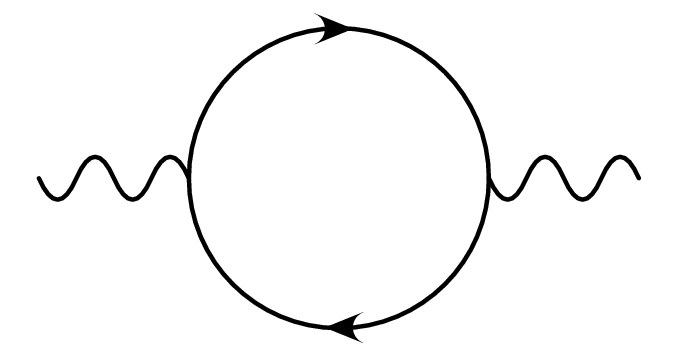}
\includegraphics[width=4.cm]{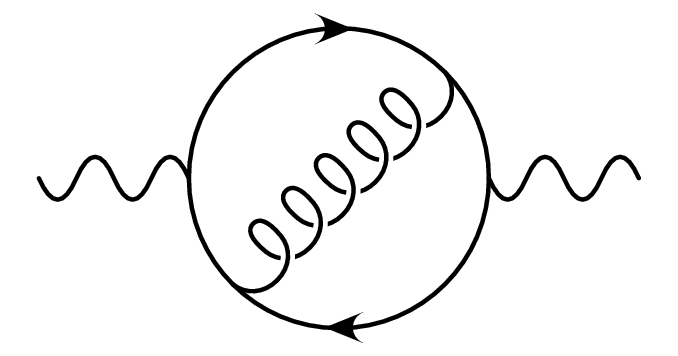}
\includegraphics[width=4.cm]{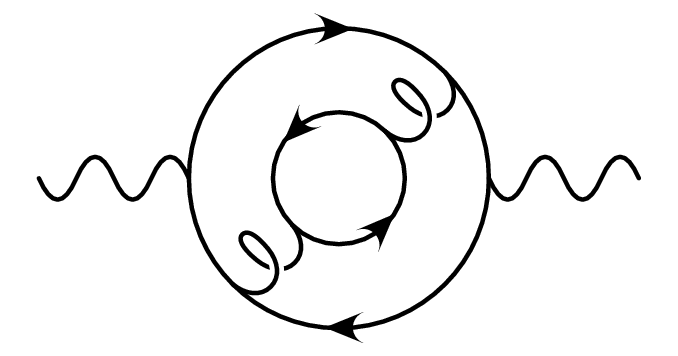}
\includegraphics[width=4.cm]{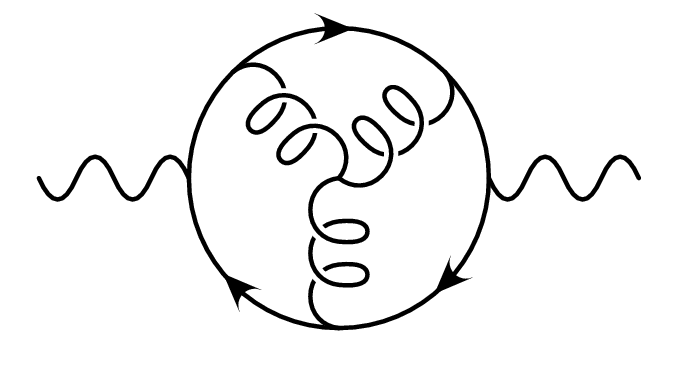}
\caption{Sample diagrams of one-,  two- and  three-loop contributions to
 the vacuum polarization of $V$.
\label{fig::samples}}
\end{figure}
The result to \nnnlo\ is well known~\cite{Chetyrkin:1979bj,Dine:1979qh,
Celmaster:1980ji,Gorishnii:1988bc,Gorishnii:1990vf},
\begin{equation}
\begin{split}
\Gamma^V_{had} =& \Gamma^V_{0,\,{\rm had}}\,{\cal F}(\amsbar,Q^2=M_V^2)
     \hskip 70pt \Gamma^V_{0,\,{\rm had}} = \frac{\alpha_V\,M_V}{3}\,N_c\sum_f\,Q_f^2\\
\sigma^{e^+\,e^-\to\ {\rm had}}(Q^2) =&
     \sigma_0^{e^+\,e^-\to\ {\rm had}}(Q^2)\,{\cal F}(\amsbar,Q^2)
     \hskip 40pt \sigma_0^{e^+\,e^-\to\ {\rm had}}(Q^2) =
       \frac{4\,\pi\,\alpha^2}{3\,Q^2}\,N_c\sum_f\,Q_f^2\\
\label{eqn:knownresult}
\end{split}
\end{equation}
and
\begin{equation}
\begin{split}
{\cal F}(\amsbar,Q^2) =\hskip-12pt&\hskip 12pt\left\{ 1 + \amsbarpi\,C_F\,\frac{3}{4}
   \LB1 + \amsbarpi\,\betabar{0}\,\ln\frac{\mu^2}{Q^2} + 
     \amsbarpi^2\Lx\betabar{1}\,\ln\frac{\mu^2}{Q^2}
    + \betabar{0}^{\,2}\,\ln^2\frac{\mu^2}{Q^2}\Rx\RB\right.\\
  & + \amsbarpi^2\LB\Lx-C_F^2\,\frac{3}{32}
   + C_F\,C_A\,\Lx\frac{123}{32}
             - \frac{11}{4}\zeta_3\Rx
   + C_F\,N_f\,\Lx-\frac{11}{16}
         + \frac{1}{2}\,\zeta_3\Rx\Rx\right.\\
  &\qquad\qquad\times\left.\Lx1 + 2\amsbarpi\,\betabar{0}
         \,\ln\frac{\mu^2}{Q^2}\Rx\RB\\
  & + \amsbarpi^3\LB-C_F^3\frac{69}{128}
 + C_F^2\,C_A\Lx-\frac{127}{64} - \frac{143}{16}\zeta_3
         + \frac{55}{4}\,\zeta_5\Rx\right.\\
  &\qquad\qquad+ C_F\,C_A^2\Lx\frac{90445}{3456} - \frac{2737}{144}\,\zeta_3
    - \frac{55}{24}\,\zeta_5\Rx\\
  &\qquad\qquad
 + C_F^2\,N_f\Lx-\frac{29}{128} + \frac{19}{8}\,\zeta_3
      - \frac{5}{2}\,\zeta_5\Rx
 + C_F\,C_A\,N_f\Lx-\frac{485}{54} + \frac{56}{9}\,\zeta_3
      + \frac{5}{12}\,\zeta_5\Rx\\
  &\left.\left.\qquad\qquad
 + C_F\,N_f^2\Lx\frac{151}{216} - \frac{19}{36}\,\zeta_3\Rx
 - \frac{1}{4}\,\pi^2\,C_F\,\betabar{0}^2\RB
  + {\cal O}\Lx\amsbarpi^4\Rx\right\}\,.
\label{eqn:knownF}
\end{split}
\end{equation}
To obtain the hadronic decay width at \lo, \nlo\ and \nnlo, I need to
compute the \qcd\ corrections to the vacuum polarization of the $V$
(photon) at $1$, $2$ and $3$ loops, respectively.  Sample diagrams are
shown in \fig{fig::samples}.

\subsection{Methods}
In each scheme, I will need to compute the vacuum polarization of $V$
and the necessary coupling renormalization constants.  As a
cross-check on the reliability of my calculational framework, I
reproduce known results on the \qcd\ $\beta$-functions and mass
anomalous dimensions to three-loop order, as well as the three-loop
\qcd\ contributions to the $\beta$-function of $V$ (where needed).

\begin{figure}[h]
a)\hbox{\includegraphics[height=1.5cm]{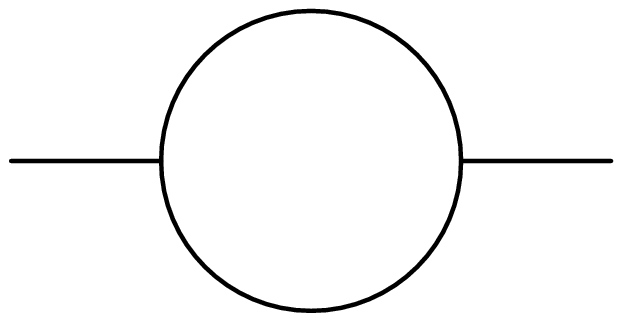}}\hskip 30pt
b)\hbox{\includegraphics[height=1.5cm]{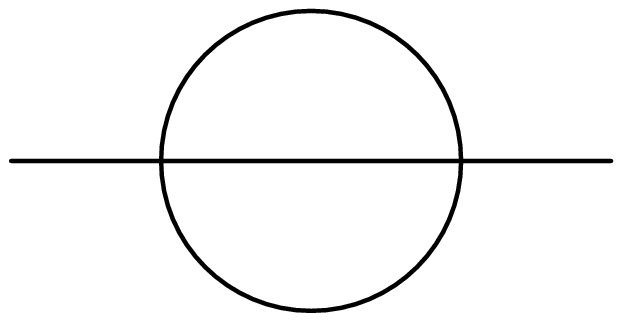}\qquad
\includegraphics[height=1.5cm]{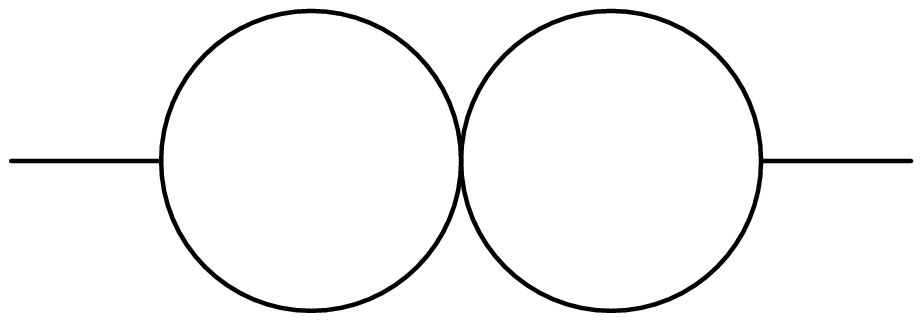}}\\[20pt]
c)\hbox{\includegraphics[height=1.5cm]{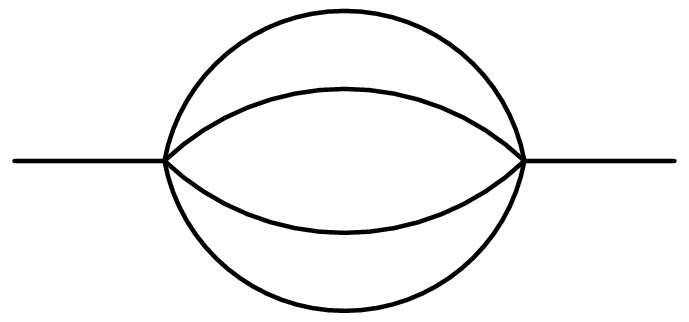}\qquad
\includegraphics[height=1.5cm]{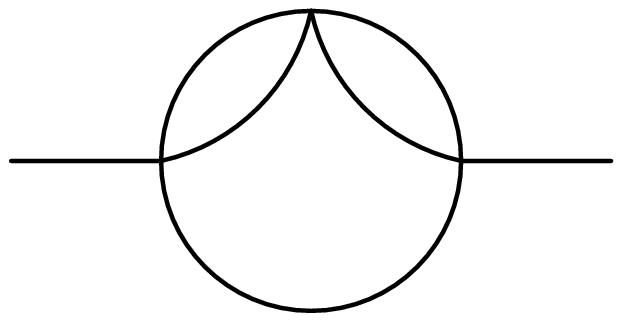}\qquad
\includegraphics[height=1.5cm]{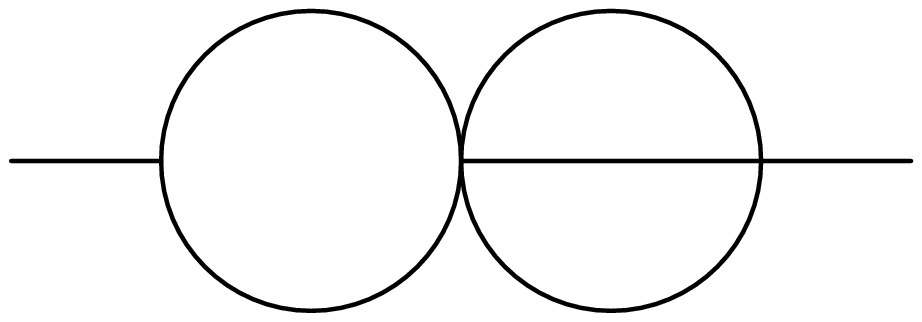}}\\[20pt]
\hskip 30pt\hbox{\includegraphics[height=1.5cm]{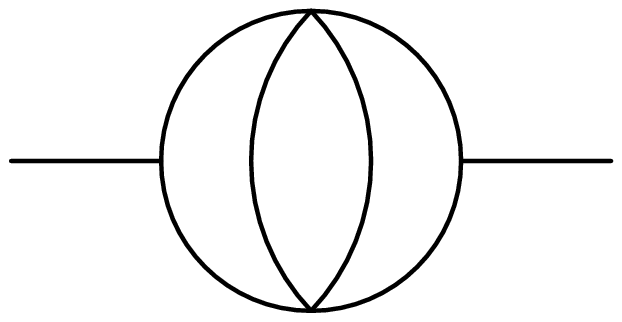}\qquad
\includegraphics[height=1.5cm]{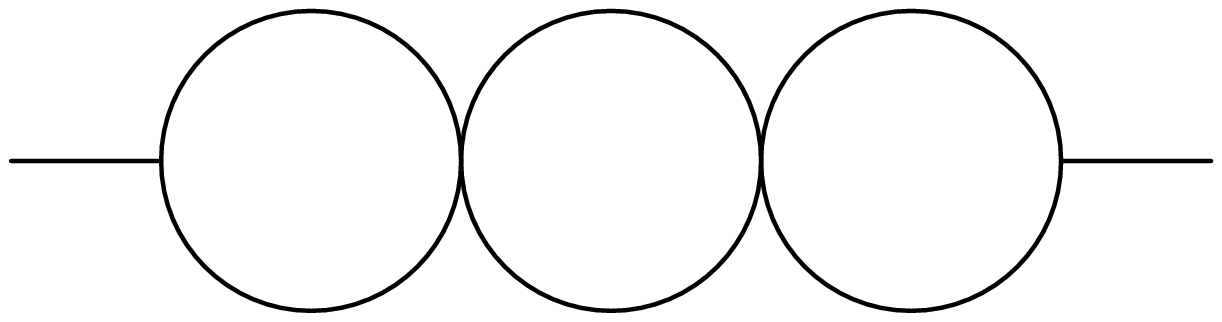}\qquad
\includegraphics[height=1.5cm]{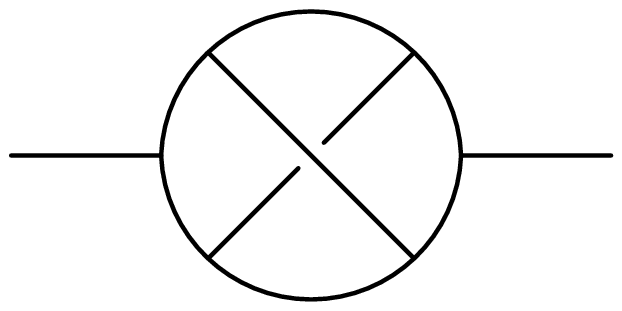}}
\caption{Master integrals for the evaluation of vacuum polarization at
  a) one loop, b) two loops and c) three loops.
\label{fig::masters}}
\end{figure}

In all calculations, I generate the contributing diagrams using
QGRAF~\cite{Nogueira:1993ex}.  The symbolic algebra program
FORM~\cite{Vermaseren:2000nd} is used to implement the Feynman rules
and perform algebraic manipulations to reduce the result to a set of
Feynman integrals to be performed and their coefficients.  The set of
Feynman integrals are then reduced to master integrals using the
program REDUZE~\cite{Studerus:2009ye}.  Using the method of
Ref.~\cite{Davydychev:1997vh}, the vertex corrections can be expressed
in terms of the same propagator integrals used to compute the vacuum
polarization and wave function renormalizations.  The complete set of
master integrals at one, two and three loops are shown in
\fig{fig::masters}.
Most of the master integrals are trivial iterated-bubble diagrams and
the others were evaluated long
ago~\cite{Chetyrkin:1980pr,Kazakov:1983ns}.  As an additional
cross-check, the integral reduction and evaluation is also performed
using the program MINCER\cite{Gorishnii:1989gt,Larin:1991fz}.

\subsection{Notation}
The various schemes that I will consider span a variety of vector
spaces, each with their own metric tensor.  To establish some level of
consistency, I will denote the metric tensor of classical
four-dimensional space-time as $\eta^{\mu\nu}$;  the metric tensor of
the $D_m$ dimensional vector space in which momentum integrals are
regularized will be denoted as $\hat{g}^{\mu\nu}$; and the metric tensor
of the largest vector space will be denoted $g^{\mu\nu}$.  Where it
does not vanish, the complement of $\hat{g}^{\mu\nu}$ will be denoted
as $\delta^{\mu\nu} = g^{\mu\nu} - \hat{g}^{\mu\nu}$.  Similarly, the
Dirac matrices $\gamma^\mu$, will be denoted $\gamma_{(4)}^\mu$ when
they are strictly four-dimensional, $\hat{\gamma}^\mu$ when they span
the $D_m$ dimensional space and $\bar{\gamma}^{\mu}$ in the space
spanned by $\delta^{\mu\nu}$.

I will now present the details of the calculation in the \cdr, \dred\
and \fdh\ schemes.

\section{Conventional Dimensional Regularization}
In the \cdr\ scheme, the calculation is quite straightforward.  The
Lagrangian and Feynman rules are just the same as for a
four-dimensional calculation, except that the Dirac matrices
$\gamma^\mu$ and the metric tensor $g^{\mu\nu}$ have been extended to
span a $D_m$ dimensional vector space.  That is,
\begin{equation}
\{\gamma^\mu,\gamma^\nu\} = 2\,g^{\mu\nu}\,,\qquad
   g^{\mu\nu}\,g_{\mu\nu} = D_m\,,\qquad
   \gamma^\mu\,\gamma_\mu = D_m\,,\qquad
   g^{\mu\nu}\equiv\hat{g}^{\mu\nu}\,.
\label{eqn:cdrdirac}
\end{equation}
The Dirac trace, $\Tr{1} = 4$, retains its standard normalization.

Although $D_m$ is given the representation $D_m = 4-2\,\ep$, the sign
of $\ep$ is not determined.  If it is taken to be positive, so that
$D_m<4$, then the Feynman integrals that one encounters are convergent
under the rules of ultraviolet power counting.  On the other hand,
infrared power counting would prefer $\ep<0 \Rightarrow D_m > 4$.  In
practice, the sign of $\ep$ does not matter and it can be used to
regularize both infrared and ultraviolet divergences.  Regardless of
the sign of $\ep$, it is important that the vector space in which
momenta take values is larger than the standard $3+1$ dimensional
space-time.  This means that the standard four-dimensional metric
tensor $\eta^{\mu\nu}$ spans a smaller space than the $D_m$
dimensional metric tensor, and the four-dimensional Dirac matrices
$\gamma^{0,1,2,3}$ form a subset of the full $\gamma^\mu$,
\begin{equation}
g^{\mu\nu}\,g_{\mu}^{\rho} = g^{\nu\rho}\,,\qquad\qquad
g^{\mu\nu}\,\eta_{\mu}^{\rho} = \eta^{\nu\rho}\,,\qquad\qquad
\eta^{\mu\nu}\,\eta_{\mu}^{\rho} = \eta^{\nu\rho}\,.
\label{eqn:cdr4dim}
\end{equation}
These considerations are of particular importance when considering
chiral objects involving $\gamma_5$ and the Levi-Civita tensor, but
will play a role in our discussion below.

Because the Dirac trace is unchanged, fermions still have exactly two
degrees of freedom in the \cdr\ scheme.  Gauge bosons, however,
acquire extra spin degrees of freedom in the $D_m$ dimensional vector
space.  The spin sum over polarization vectors in a physical (axial)
gauge takes the form
\begin{equation}
-g_{\mu\nu}\,\sum_{\lambda} \epsilon^{*\,\mu}(k,\lambda)\,\epsilon^{\nu}(k,\lambda)
   = g_{\mu\nu}\,\Lx g^{\mu\nu} - \frac{k^\mu\,n^\nu +
   n^\mu\,k^\nu}{k\cdot n}\Rx = D_m-2 =  2-2\,\ep\,,
\label{eq::cdrspinsum}
\end{equation}
where $n$ is the axial gauge reference vector.  For massive vector
bosons, the spin sum becomes
\begin{equation}
-g_{\mu\nu}\,\sum_{\lambda} \epsilon^{*\,\mu}(k,\lambda)\,\epsilon^{\nu}(k,\lambda)
   = g_{\mu\nu}\,\Lx g^{\mu\nu} - \frac{k^\mu\,k^\nu}{M^2}\Rx = D_m-1 =  3-2\,\ep\,,
\label{eq::cdrmassivespinsum}
\end{equation}

\subsection{Renormalization}
The renormalization constants in the \cdr\ scheme are defined as
\begin{equation}
\begin{split}
  \Gamma^{(B)}_{AAA}&= Z_{1}\Gamma_{AAA}\,,\qquad
  \psi^{(B)\,i}_f = Z^{\frac{1}{2}}_2\,\psi^{i}_f\,,\qquad
  A^{(B)\, a}_\mu = Z^{\frac{1}{2}}_3\,A^{a}_\mu\\
  \Gamma^{(B)}_{c\overline{c}A}&= \widetilde{Z}_1
     \Gamma_{q\overline{q}A}\,,\qquad\
  c^{(B)\,a} = \widetilde{Z}^{\frac{1}{2}}_3\,c^a\,,\qquad\ \
  \overline{c}^{(B)\,a} = \widetilde{Z}^{\frac{1}{2}}_3\,\overline{c}^a\,,\\
  \Gamma^{(B)}_{q\overline{q}A}&= Z_{1\,F}\Gamma_{q\overline{q}A}\,,\qquad
  \xi^{(B)} = \xi\,Z_3\,,
\end{split}
\end{equation}
where $\Gamma_{abc}$ represents the vertex function involving fields
$a$, $b$ and $c$.

Although we treat the quark fields as massless, we can compute the
mass anomalous dimension by introducing a fictitious scalar particle
$\phi$ and computing the $\beta$-function of its Yukawa coupling to
the quarks.  The equivalence is clear from the standard model, where
the Higgs Yukawa coupling and the fermion mass are proportional at
leading electroweak order and must behave the same under \qcd\
renormalization.  For this purpose, I introduce one more
renormalization constant, $\Gamma^{(B)}_{q\overline{q}\phi}=
Z_{1\,\phi}\Gamma_{q\overline{q}\phi}$.  One can introduce a
wave function renormalization for $\phi$, $Z_{3\,\phi}$, but it will
not contribute because $Z_{3\,\phi} = 1 + {\cal O}(\alpha_\phi)$.
Note also that I do not need to compute the \qcd\ corrections to the
$\beta$-function for $\alpha_V$, which will start at order
$\alpha_V^2$ because of the Ward Identity.

In the $\overline{\rm MS}$ scheme, the couplings renormalize as
\begin{equation}
\begin{split}
\asbare &=\Lx\frac{\mu^2\,e^{\gamma_E}}{4\,\pi}\Rx^\ep\,
    Z_{\amsbar}\,\amsbar\,,\qquad Z_{\amsbar} = \frac{Z_1^2}{Z_3^3}
  = \frac{Z_{1\,F}^2}{Z_2^2\,Z_3}
    = \frac{\widetilde{Z}_1^2}{\widetilde{Z}_3^2\,Z_3}\\
\aphibare &=\Lx\frac{\mu^2\,e^{\gamma_E}}{4\,\pi}\Rx^\ep\,
    Z_{\aphimsbar}\,\aphimsbar\,,\qquad Z_{\aphimsbar}
    = \frac{Z_{1\,\phi}^2}{Z_2^2\,Z_{3\,\phi}}
\label{eqn::cdrrenorm}
\end{split}
\end{equation}

The structure of the renormalization constants $Z_{\amsbar}$ and
$Z_{\aphimsbar}$ is determined entirely by their lowest order
($1/\ep$) poles, which in turn define the $\beta$-functions.

\begin{equation}
\begin{split}
\betabar{}(\amsbar) = \mu^2\frac{d}{d\,\mu^2}\frac{\amsbar}{\pi}
   &= -\ep\frac{\amsbar}{\pi}\Lx1 + \frac{\amsbar}{Z_{\amsbar}}
    \frac{\partial Z_{\amsbar}}{\partial\amsbar}\Rx^{-1}\\
   &= -\ep\frac{\amsbar}{\pi} - \sum_{n=0}^\infty\,\betabar{n}\,\amsbarpi^{n+2}\\
\betabar{\phi}(\amsbar) = \mu^2\frac{d}{d\,\mu^2}\frac{\aphimsbar}{\pi}
   &= -\Lx\ep\frac{\aphimsbar}{\pi} + \frac{\aphimsbar}{Z_{\aphimsbar}}
    \frac{\partial Z_{\aphimsbar}}{\partial\amsbar}
    \,\betabar{}(\amsbar)\Rx\Lx1 + \frac{\aphimsbar}{Z_{\aphimsbar}}
    \frac{\partial Z_{\aphimsbar}}{\partial\aphimsbar}\Rx^{-1}\\
   &= -\frac{\aphimsbar}{\pi}\Lx\ep + \sum_{n=0}^\infty
     \,\betabar{\phi\,,n}\,\amsbarpi^{n+1}\Rx\\
\label{eqn:cdrbetadef}
\end{split}
\end{equation}
The mass anomalous dimension,
\begin{equation}
\gammabar{}(\amsbar) =
    \frac{\mu^2}{m^{\msbar}}\frac{d}{d\mu^2}m^{\msbar}
   = \sum_{n=0}^\infty -\gammabar{n}\amsbarpi^{n+1}
\end{equation}
is defined in terms of $m$, rather than $m^2$, with the result that
$\gammabar{n} = \frac{1}{2}\betabar{\phi\,,n}$.  The results for
$\betabar{n}$ and $\gammabar{n}$ through three loops are given in
\app{sec:cdrrenorm}.

\subsection{Vacuum polarization in the \cdr\ scheme}
The imaginary part of the unrenormalized vacuum polarization tensor in
the \cdr\ scheme is
\begin{equation}
\begin{split}
\Im\LB\left.\Pi^{(B)}_{\mu\nu}(Q)\right|_{\cdr}\RB &=
   \frac{ -Q^2\,g_{\mu\nu} + Q_{\mu}Q_{\nu}}{3}
   \avbare\,N_c\,\sum_f\,Q_f^2\Lx\frac{4\,\pi}{Q^2\,e^{\gamma_E}}\Rx^\ep
          \left\{ \vphantom{\asbarepi}\right.\\
    & \hskip-50pt
     1 + \asbarepi\,\Lx\frac{4\,\pi}{Q^2\,e^{\gamma_E}}\Rx^{\ep}
       C_F\,\LB\frac{3}{4}
      + \ep\Lx\frac{55}{8} - 6\,\zeta_3\Rx
      + \ep^2\,\Lx\frac{1711}{48} - \frac{15}{4}\,\zeta_2 - 19\,\zeta_3
           - 9\,\zeta_4\Rx
            + {\cal O}(\ep^3)\RB\\
    & \hskip-50pt
     +\asbarepi^2\Lx\frac{4\,\pi}{Q^2\,e^{\gamma_E}}\Rx^{2\,\ep}\LB
       \frac{1}{\ep}\Lx\,\frac{11}{16}C_F\,C_A
             - \frac{1}{8}C_F\,N_f\Rx\right.\\
    &\hskip -30pt    - \frac{3}{32}\,C_F^2 + C_F\,C_A\Lx\frac{487}{48}
             - \frac{33}{4}\zeta_3\Rx
             + C_F\,N_f\Lx - \frac{11}{6} + \frac{3}{2}\,\zeta_3\Rx\\
    &\hskip -30pt +\ep\Lx C_F^2\Lx - \frac{143}{32} - \frac{111}{8}\,\zeta_3
                    + \frac{45}{2}\,\zeta_5\Rx
        + C_F\,C_A\Lx \frac{50339}{576} - \frac{231}{32}\,\zeta_2 - \frac{109}{2}\,\zeta_3
                    - \frac{99}{8}\,\zeta_4 -
                 \frac{15}{4}\,\zeta_5\Rx\right.\\
    &\hskip -30pt\left.\left.\left.
        + C_F\,N_f\Lx - \frac{4417}{288} + \frac{21}{16}\,\zeta_2
                    + \frac{19}{2}\,\zeta_3 + \frac{9}{4}\,\zeta_4\Rx\Rx
        + {\cal O}(\ep^2)\RB  + {\cal O}\Lx\asbarepi^3\Rx\right\}\,.
\label{eqn:impi0cdr}
\end{split}
\end{equation}
Upon renormalizing the \qcd\ coupling according to
\eqn{eqn::cdrrenorm}, setting $\avbare\to\av\Lx\frac{\mu^2\,e^{\gamma_E}}{4\,\pi}\Rx^\ep$, and dropping terms of
order $(\ep)$, I obtain
\begin{equation}
\begin{split}
\Im\LB\left.\Pi_{\mu\nu}(Q)\right|_{\cdr}\RB &=
   \frac{ -Q^2\,g_{\mu\nu} + Q_{\mu}Q_{\nu}}{3}
   \av\,N_c\,\sum_f\,Q_f^2\,\left\{  1 + \amsbarpi\, C_F\,\frac{3}{4}\LB1
       + \amsbarpi\,\betabar{0}\,\ln\frac{\mu^2}{Q^2}\RB\right.\\
    & \hskip-45pt\left.
     +\amsbarpi^2\,\LB -C_F^2\,\frac{3}{32}
   + C_F\,C_A\,\Lx\frac{123}{32}
             - \frac{11}{4}\zeta_3\Rx
   + C_F\,N_f\,\Lx-\frac{11}{16}
         + \frac{1}{2}\,\zeta_3\Rx
       \RB  + {\cal O}\Lx\amsbarpi^3\Rx\right\}\,.
\label{eqn:impicdr}
\end{split}
\end{equation}
In this way of performing the calculation, all of the \qcd\ states
that appear are internal states, so the \hv\ scheme gives exactly the
same result.

\subsection{Total Decay rate and annihilation cross section in the
  \cdr\ scheme}
\label{sec:cdrdecann}

The decay rate and the annihilation cross section are
determined by computing the imaginary part of the forward scattering
amplitude.  For the decay rate, this means attaching the polarization
vector $\ep^{\mu}(Q,\lambda)$ and its conjugate
$\ep^{\nu}(Q,\lambda)^{*}$ ($Q^2 = M_V^2$) and averaging over the
spins,
\begin{equation}
\Gamma^{\cdr}_{V\to\ {\rm hadrons}} = \frac{1}{M_V}\frac{1}{N_{\rm spins}}
    \sum_\lambda \ep^{\mu}(Q,\lambda)\,\Im\LB\left.\Pi_{\mu\nu}(Q)
    \right|_{\cdr}\RB \,\ep^{\nu}(Q,\lambda)^{*}\,,
\label{eqn:cdrdecay}
\end{equation}
where
\begin{equation}
\frac{1}{N_{\rm spins}}\sum_\lambda \ep^{\mu}(Q,\lambda)\,
    \ep^{\nu}(Q,\lambda)^{*}
   = \frac{1}{N_{\rm spins}}\Lx-g^{\mu\nu} + \frac{Q^\mu\,Q^\nu}{M_V^2}\Rx\,.
\label{eqn:cdrspinav}
\end{equation}

Notice that because the imaginary part of the vacuum polarization
tensor is finite, it does not matter whether the spin sum is taken in
$D_m = 4 - 2\,\ep$ dimensions as in the \cdr\ scheme or in four
dimensions as in the \hv\ scheme as the difference is of order $\ep$.
The result is
\begin{equation}
\begin{split}
\Gamma^{\cdr}_{V\to\ {\rm hadrons}} =&
    \frac{\av\,M_V}{3}\,N_c\,\sum_f\,Q_f^2\,
       \left\{  1 + \amsbarpi\, C_F\,\frac{3}{4}\LB1
       + \amsbarpi\,\betabar{0}\,\ln\frac{\mu^2}{Q^2}\RB\right.\\
    & \hskip-45pt\left.
     +\amsbarpi^2\,\LB -C_F^2\,\frac{3}{32}
   + C_F\,C_A\,\Lx\frac{123}{32}
             - \frac{11}{4}\zeta_3\Rx
   + C_F\,N_f\,\Lx-\frac{11}{16}
         + \frac{1}{2}\,\zeta_3\Rx
       \RB  + {\cal O}\Lx\amsbarpi^3\Rx\right\}\,,
\label{eqn:gamcdrverify}
\end{split}
\end{equation}
in agreement with \eqns{eqn:knownresult}{eqn:knownF}.

For the annihilation cross section $\sigma_{e^+\,e^-\to\ {\rm
    hadrons}}$, one attaches fermion bilinears to each end of the
vacuum polarization tensor and averages over the spins.
\begin{equation}
\sigma^{\cdr}_{e^+\,e^-\to\ {\rm hadrons}} = \frac{2}{Q^2}\frac{e^2}{4}
   \sum_{\lambda\,\lambda^{'}} \frac{\mtrxelm{\overline{v}(p_{e^+},\lambda)}
   {\gamma^{\mu}}{u(p_{e^-},\lambda^{'})}}{Q^2}\Im\LB\left.\Pi_{\mu\nu}(Q)
    \right|_{\cdr,\,\av\to\alpha}\RB\frac{\mtrxelm{\overline{u}(p_{e^-},\lambda^{'})}
    {\gamma^{\nu}}{v(p_{e^+},\lambda)}}{Q^2}\,.
\label{eqn:cdrannav}
\end{equation}
Because this is a forward scattering amplitude, the spinor bilinears
can be combined into a trace,
\begin{equation}
   \frac{1}{2} \sum_{\lambda\,\lambda^{'}} \mtrxelm{
    \overline{v}(p_{e^+},\lambda)}{\gamma^{\mu}}{u(p_{e^-},\lambda^{'})}
    \mtrxelm{\overline{u}(p_{e^-},\lambda^{'})} {\gamma^{\nu}}
    {v(p_{e^+},\lambda)} = \frac{1}{2}\Tr{\slashed{p}_{e^+}\,
    \gamma^{\mu}\slashed{p}_{e^-}\,\gamma^{\nu}}
   = \Lx-Q^2\,g^{\mu\,\nu} + Q^\mu\,Q^\nu\Rx\,,
\end{equation}
where the last identification results from the fact that $Q^\mu =
p_{e^-}^{\mu} + p_{e^+}^{\mu}$,\quad $p_{e^-}\cdot\,Q =
p_{e^+}\cdot\,Q = Q^2/2$.  The result is
\begin{equation}
\begin{split}
\sigma^{\cdr}_{e^+\,e^-\to\ {\rm hadrons}} =&
    \frac{4\pi\,\alpha^2}{3\,Q^2}\,N_c\,\sum_f\,Q_f^2\,
       \left\{  1 + \amsbarpi\, C_F\,\frac{3}{4}\LB1
       + \amsbarpi\,\betabar{0}\,\ln\frac{\mu^2}{Q^2}\RB\right.\\
    & \hskip-45pt\left.
     +\amsbarpi^2\,\LB -C_F^2\,\frac{3}{32}
   + C_F\,C_A\,\Lx\frac{123}{32}
             - \frac{11}{4}\zeta_3\Rx
   + C_F\,N_f\,\Lx-\frac{11}{16}
         + \frac{1}{2}\,\zeta_3\Rx
       \RB  + {\cal O}\Lx\amsbarpi^3\Rx\right\}\,,
\label{eqn:sigcdrverify}
\end{split}
\end{equation}
again in agreement with \eqns{eqn:knownresult}{eqn:knownF}.

Thus, I have established that I can reproduce the known results in the
\cdr\ scheme through three-loop order, which is a strong check on my
computational framework.

\section{Dimensional Reduction}
In dimensional reduction, one starts from standard four-dimensional
space-time and compactifies to a {\it smaller} vector space of
dimension $D_m = 4 - 2\,\ep < 4$ in which momenta take values.  The
particles in the spectrum, however, retain the spin degrees of freedom
of four dimensions.  That is, both fermions and gauge bosons still
have two degrees of freedom.  This is by design, of course, since it
is required by supersymmetry.  All Dirac algebra can be treated as
four-dimensional.  However, now the four-dimensional metric tensor
$\eta^{\mu\nu}$ spans a larger space than the $D_m$ dimensional metric
$\hat{g}^{\mu\nu}$ that might arise from tensor momentum integrals,
\begin{equation}
\hat{g}^{\mu\nu}\,\eta_{\mu}^{\rho} = \hat{g}^{\nu\rho}\,.
\label{eqn::metricdred}
\end{equation}

There is also a very serious consequence of the fact that the $D_m$
dimensional vector space is smaller than four-dimensional space-time.
The Ward Identity only applies to the $D_m$ dimensional vector space.
This means that the $2\,\ep$ spin degrees of freedom that are not
protected by the Ward Identity must renormalize differently than the
$2-2\,\ep$ degrees of freedom that are protected.  In supersymmetric
theories, the supersymmetry provides the missing Ward Identity which
demands that the $2\,\ep$ spin degrees of freedom be treated as
gauge bosons.  In nonsupersymmetric theories, however, they must be
considered to be distinct particles, with distinct couplings and
renormalization properties.  It is common to refer to these extra
degrees of freedom as ``$\ep$-scalars'' or as ``evanescent'' degrees
of freedom.

Once the evanescent degrees of freedom (which I will label
$A_e^{a\,\tilde\mu}$, to distinguish them from the gluons,
$A^{a\,\mu}$) are recognized as independent particles, it is apparent
that their couplings are also independent, not only of the \qcd\
coupling, but of one another.  That is, the coupling $g_e$ of the
evanescent gluons to the quarks is not only distinct from $g$, the
coupling of \qcd, but is also distinct from $\lambda_i$, the quartic
couplings of the evanescent gluons to themselves.  (The quartic gauge
coupling of \qcd\ splits into three independent quartic couplings of
the evanescent gluons.)  Note that the massive vector boson $V^\mu$
also has evanescent degrees of freedom, $V_e^{\tilde\mu}$, which
couple to quarks with strength $g_{Ve}$.

Thus, the Lagrangian in the \dred\ scheme becomes:
\begin{equation}
\begin{split}
{\cal L} =& - \frac{1}{2}A^{a}_{\mu}\Lx\partial^{\mu}\partial^{\nu}
  (1-\xi^{-1}) - \hat{g}^{\mu\nu}\Box\Rx A^{a}_{\nu}
 - g\,f^{abc}(\partial^\mu\,A^{a\,\nu})A^b_\mu\,A^c_\nu
 - \frac{g^2}{4}f^{abc}\,f^{ade}\,A^{b\,\mu}\,A^{c\,\nu}\,A^d_\mu\,A^e_\nu\\
& + i\sum_f\,
\overline{\psi}_f^{i}\Lx\delta_{ij}\slashed{\partial}
    -i\,g\,t^{a}_{ij}\slashed{A}^a
    -i\,g_V\,Q_f\slashed{V}\Rx\,\psi_f^{j}
  - \overline{c}^{a}\Box\,c^a
    + g\,f^{abc}\Lx\partial_\mu\,\overline{c}^a\Rx\,A^{b\,\mu}\,c^c\\
& + \frac{1}{2}A_{e\,\tilde\mu}^{a}\Box\ A_e^{a\,\tilde\mu}
 - g\,f^{abc}(\partial^\mu\,A_e^{a\,\tilde\nu})A^b_\mu\,A^c_{e\,\tilde\nu}
 + \frac{g^2}{2}f^{abc}\,f^{adf}\,A^{b\,\mu}\,A_e^{c\,\tilde\nu}\,
    A^d_\mu\,A_{e\,\tilde\nu}^f
 - \frac{1}{4}\sum_i \lambda_i\,H_i^{bcdf}\,A_e^{b\,\tilde\mu}\,A_e^{c\,\tilde\nu}\,
    A_{e\,\tilde\mu}^d\,A_{e\,\tilde\nu}^f\\
& + \sum_f\,\overline{\psi}_f^{i}\Lx g_e\,t^{a}_{ij}\slashed{A}_e^a
    + g_{Ve}\,Q_f\slashed{V}_e\Rx\,\psi_f^{j}\,.
\label{eqn::dredlagrange}
\end{split}
\end{equation}
As mentioned above, the quartic coupling of the evanescent gluons
splits into three terms, which mix under renormalization.  One can
choose the tensors $H_i^{bcde}$ to be~\cite{Harlander:2006xq}
\begin{equation}
\begin{split}
H_1^{bcde} =& \frac{1}{2}\Lx f^{abc}\,f^{ade} + f^{abe}\,f^{adc}\Rx\\
H_2^{bcde} =& \delta^{bc}\delta^{de} + \delta^{bd}\delta^{ce}
                      + \delta^{be}\delta^{cd}\\
H_3^{bcde} =& \frac{1}{2}\Lx\delta^{bc}\delta^{de} + \delta^{be}\delta^{cd}\Rx
                      - \delta^{bd}\delta^{ce}\,,
\label{eqn::dredquartic}
\end{split}
\end{equation}
Although the quartic couplings enter the $\beta$-functions and
anomalous dimension at three loops and are essential to the
renormalization program, they do not explicitly contribute to the
calculation at hand.

Now that the correct spectrum has been identified, one must carefully
consider the renormalization program.  The na\"ive application of the
principle of minimal subtraction leads to the violation of
unitarity~\cite{vanDamme:1984ig}.  Because the contributions of
evanescent states and couplings to scattering amplitudes are weighted
by a factor $\ep$, the leading one-loop contribution is finite and
therefore not subtracted.  As one proceeds to higher orders, there is
a mismatch among the counterterms such that the renormalization
program fails to remove all of the ultraviolet singularities.

A successful renormalization program for the \dred\
scheme~\cite{Jack:1993ws,Jack:1994bn} applies the principle of minimal
subtraction to the evanescent Green functions (that is, Green
functions with external evanescent states) themselves.  At each order,
the renormalization scheme renders the evanescent Green functions
finite.  Since evanescent Green functions enter into the scattering
amplitudes of physical particles at order $\ep$ and they are rendered
finite by renormalization, they never contribute to physical
scattering amplitudes.

The evanescent coupling still contributes to Green functions with only
physical external states, but the contribution is rendered finite by
the prescribed renormalization
program~\cite{Jack:1993ws,Jack:1994bn,Harlander:2006rj,Harlander:2006xq}.
Because the evanescent coupling, $\alpha_e$ renormalizes differently
than the gauge coupling $\alpha_s$, the two cannot be identified, even
at the end of the calculation.  One can choose a renormalization point
where the two coincide, but they evolve differently under
renormalization group transformations and their values will diverge as
one moves away from the renormalization point.

Still, the evanescent coupling is essentially a fictitious quantity
and one finds that if one computes a physical quantity in the \dred\
scheme and then converts the running couplings of the \dred\ scheme to
those of a scheme such as \cdr\ that has no evanescent couplings, the
factors of $\alpha_e$ drop
out~\cite{Harlander:2006rj,Harlander:2006xq}.

\subsection{Renormalization}
The renormalization constants in the \dred\ scheme are defined as
\begin{equation}
\begin{split}
  \Gamma^{(B)}_{AAA}&= Z_{1}\Gamma_{AAA}\,,\qquad
  \psi^{(B)\,i}_f = Z^{\frac{1}{2}}_2\,\psi^{i}_f\,,\qquad
  A^{(B)\, a}_\mu = Z^{\frac{1}{2}}_3\,A^{a}_\mu\\
  \Gamma^{(B)}_{c\overline{c}A}&= \widetilde{Z}_1
     \Gamma_{q\overline{q}A}\,,\qquad\
  c^{(B)\,a} = \widetilde{Z}^{\frac{1}{2}}_3\,c^a\,,\qquad\ \
  \overline{c}^{(B)\,a} = \widetilde{Z}^{\frac{1}{2}}_3\,\overline{c}^a\,,\\
  \Gamma^{(B)}_{q\overline{q}A}&= Z_{1\,F}\Gamma_{q\overline{q}A}\,,\qquad
  \xi^{(B)} = \xi\,Z_3\,,\\
  \Gamma^{(B)}_{q\overline{q}e}&=Z_{1\,e}\Gamma_{q\overline{q}e}\,,\qquad 
  A^{(B)\,a}_{e\, \mu} = Z^{\frac{1}{2}}_{3\,e}\,A^{a}_{e\,\mu}\,,\qquad 
  \Gamma^{(B)\,i}_{eeee} = Z^{i}_{1\,eeee}\,\Gamma^{i}_{eeee}\,,\\
  \Gamma^{(B)}_{q\overline{q}V_e}&=Z_{1\,Ve}\Gamma_{q\overline{q}V_e}\,,\qquad 
  V^{(B)}_{e\, \mu} = Z^{\frac{1}{2}}_{3\,Ve}\,V_{e\,\mu}\,.
\label{eqn:dredrenormconst}
\end{split}
\end{equation}
In addition, I again introduce the fictitious scalar that allows me to
compute the mass anomalous dimension for massless quarks.  Note that
while the Ward Identity protects $\alpha_V$ from leading \qcd\
corrections, it does not protect $\alpha_{Ve}$.  That is why I need to
introduce renormalization constants for the vertex and wave-function and
why I need to compute the $\beta$-function of $\alpha_{Ve}$.

In the $\drbar$ scheme (modified minimal subtraction in the \dred\
scheme), the couplings renormalize as
\begin{equation}
\begin{split}
\asbare &=\Lx\frac{\mu^2\,e^{\gamma_E}}{4\,\pi}\Rx^\ep\,
    Z_{\adrbar}\,\adrbar\,,\qquad Z_{\adrbar} = \frac{Z_1^2}{Z_3^3}
  = \frac{Z_{1\,F}^2}{Z_2^2\,Z_3}
    = \frac{\widetilde{Z}_1^2}{\widetilde{Z}_3^2\,Z_3}\,,\\
\aebare &=\Lx\frac{\mu^2\,e^{\gamma_E}}{4\,\pi}\Rx^\ep\,
    Z_{\aedrbar}\,\aedrbar\,,\qquad Z_{\aedrbar}
    = \frac{Z_{1\,e}^2}{Z_2^2\,Z_{3\,e}}\,,\\
\avebare &=\Lx\frac{\mu^2\,e^{\gamma_E}}{4\,\pi}\Rx^\ep\,
    Z_{\avedrbar}\,\avedrbar\,,\qquad Z_{\avedrbar}
  = \frac{Z_{1\,Ve}^2}{Z_2^2\,Z_{3\,Ve}}\,,\\\
\aphibare &=\Lx\frac{\mu^2\,e^{\gamma_E}}{4\,\pi}\Rx^\ep\,
    Z_{\aphidrbar}\,\aphidrbar\,,\qquad Z_{\aphidrbar}
    = \frac{Z_{1\,\phi}^2}{Z_2^2\,Z_{3\,\phi}}\,.
\label{eqn:dredrenorm}
\end{split}
\end{equation}

and the $\beta$-functions are given by
\begin{equation}
\begin{split}
\betadred{} = \mu^2\frac{d}{d\,\mu^2}\frac{\adrbar}{\pi}
   &= -\Lx\ep\frac{\adrbar}{\pi}
   +\frac{\adrbar}{Z_{\adrbar}}\frac{\partial
      Z_{\adrbar}}{\partial\aedrbar}\,\betadred{e}
   +\frac{\adrbar}{Z_{\adrbar}}\frac{\partial
      Z_{\adrbar}}{\partial\etadrbar{i}}\,\betadred{\eta_i}\Rx 
    \Lx1 + \frac{\adrbar}{Z_{\adrbar}}
    \frac{\partial Z_{\adrbar}}{\partial\adrbar}\Rx^{-1}\\
   &= -\ep\frac{\adrbar}{\pi} - \sum_{i,j,k,l,m}\,\betadred{ijklm}
    \,\adrbarpi^{i}\,\aedrbarpi^{j}\,\etadrbarpi{1}^{k}
    \,\etadrbarpi{2}^{l}\,\etadrbarpi{3}^{m}\\
\betadred{e} = \mu^2\frac{d}{d\,\mu^2}\frac{\aedrbar}{\pi}
   &= -\Lx\ep\frac{\aedrbar}{\pi}
   +\frac{\aedrbar}{Z_{\aedrbar}}\frac{\partial
      Z_{\aedrbar}}{\partial\adrbar}\,\betadred{}
   +\frac{\aedrbar}{Z_{\aedrbar}}\frac{\partial
      Z_{\aedrbar}}{\partial\etadrbar{i}}\,\betadred{\eta_i}\Rx 
    \Lx1 + \frac{\aedrbar}{Z_{\aedrbar}}
    \frac{\partial Z_{\aedrbar}}{\partial\aedrbar}\Rx^{-1}\\
   &= -\ep\frac{\aedrbar}{\pi} - \sum_{i,j,k,l,m}\,\betadred{e,\,ijklm}
    \,\adrbarpi^{i}\,\aedrbarpi^{j}\,\etadrbarpi{1}^{k}
    \,\etadrbarpi{2}^{l}\,\etadrbarpi{3}^{m}\\
\betadred{Ve} = \mu^2\frac{d}{d\,\mu^2}\frac{\avedrbar}{\pi}
   &= -\Lx\ep\frac{\avedrbar}{\pi}
   +\frac{\avedrbar}{Z_{\avedrbar}}\frac{\partial
      Z_{\avedrbar}}{\partial\adrbar}\,\betadred{}
   +\frac{\avedrbar}{Z_{\avedrbar}}\frac{\partial
      Z_{\avedrbar}}{\partial\aedrbar}\,\betadred{e}
   +\frac{\avedrbar}{Z_{\avedrbar}}\frac{\partial
      Z_{\avedrbar}}{\partial\etadrbar{i}}\,\betadred{\eta_i}\Rx\\
   &\hskip 100pt \times
    \Lx1 + \frac{\avedrbar}{Z_{\avedrbar}}
    \frac{\partial Z_{\avedrbar}}{\partial\avedrbar}\Rx^{-1}\\
   &= -\frac{\avedrbar}{\pi}\Lx\ep + \sum_{i,j,k,l,m}\,\betadred{Ve,\,ijklm}
    \,\adrbarpi^{i}\,\aedrbarpi^{j}\,\etadrbarpi{1}^{k}
    \,\etadrbarpi{2}^{l}\,\etadrbarpi{3}^{m}\Rx\\
\betadred{\phi} = \mu^2\frac{d}{d\,\mu^2}\frac{\aphidrbar}{\pi}
   &= -\Lx\ep\frac{\aphidrbar}{\pi} + \frac{\aphidrbar}{Z_{\aphidrbar}}
    \frac{\partial Z_{\aphidrbar}}{\partial\adrbar}
    \,\betadred{} + \frac{\aphidrbar}{Z_{\aphidrbar}}
    \frac{\partial Z_{\aphidrbar}}{\partial\aedrbar}
    \,\betadred{e} + \frac{\aphidrbar}{Z_{\aphidrbar}}
    \frac{\partial Z_{\aphidrbar}}{\partial\etadrbar{i}}
    \,\betadred{\eta_i}\Rx\\
   &\hskip 100pt \times\Lx1 + \frac{\aphidrbar}{Z_{\aphidrbar}}
    \frac{\partial Z_{\aphidrbar}}{\partial\aphidrbar}\Rx^{-1}\\
   &= -\frac{\aphidrbar}{\pi}\Lx\ep + \sum_{i,j,k,l,m}\,\betadred{\phi,\,ijklm}
    \,\adrbarpi^{i}\,\aedrbarpi^{j}\,\etadrbarpi{1}^{k}
    \,\etadrbarpi{2}^{l}\,\etadrbarpi{3}^{m}\Rx\\
\label{eqn:dredbetadef}
\end{split}
\end{equation}

Through three-loop order, the $\eta_i$ do not contribute to the \qcd\
$\beta$-function, $\betadred{}$, nor to the vacuum polarization of $V$
(or $V_e$).  To three-loop order, I find agreement with known
results~\cite{Harlander:2006rj,Harlander:2006xq} and derive new
results for the $\beta$-function of $\alpha_{Ve}$.  The coefficients
of the $\beta$-functions and anomalous dimensions are given in
\app{sec:dredrenorm}.

By comparing $\betavedred{}{}$ and $\gammadred{}$ in
\eqns{eqn:dredgamma}{eqn:dredbetave}, we see that the term
``$\ep$-scalar'' is a misnomer.  If the evanescent part of $V$ were a
true scalar, its $\beta$-function would coincide (but for a factor of
$2$) with the mass anomalous dimension.  The pure $\adrbar$ terms do
coincide, because there is no nonvanishing contraction of the Lorentz
indices of the evanescent $V$ and those of the gluons.  Because there
are contractions between the Lorentz indices of the evanescent $V$ and
those of the evanescent gluons, however, terms involving $\aedrbar$ do
not agree.

Calculations in the \dred\ scheme naturally produce results in terms
of $\adrbar$ while the standard result has been expressed in terms of
$\amsbar$.  One can always convert one renormalized coupling to
another.  The rule for converting $\adrbar\to\amsbar$, derived in
Refs.~\cite{Kunszt:1993sd,Harlander:2006rj}, is
\begin{equation}
  \adrbar = \amsbar\LB1 + \amsbarpi\frac{C_A}{12}
  + \amsbarpi^2\frac{11}{72}C_A^2
  - \amsbarpi\aedrbarpi\frac{C_F\,N_f}{16}
  + \ldots\RB
\label{eqn:drbartomsbar}
\end{equation}
When the result is expressed in terms of $\amsbar$, all $\aedrbar$
terms drop out.

\subsection{Vacuum polarization in the \dred\ scheme}
In the \dred\ scheme, there are two independent transverse vacuum polarization tensors,
\begin{equation}
  \Im\LB\left.\Pi^{(B)}_{\mu\nu}(Q)\right|_{\dred}\RB =
    \frac{-Q^2\,\hat{g}_{\mu\nu} + Q_{\mu}Q_{\nu}}{3}\,
       \Im\LB\left.\Pi^{(B)}_{A}(Q)\right|_{\dred}\RB
   - Q^2\,\frac{\delta_{\mu\nu}}{2\,\ep}\,\Im\LB\left.\Pi^{(B)}_{B}(Q)\right|_{\dred}\RB\,,
\end{equation}
where
\begin{equation}
\begin{split}
\Im\LB\left.\Pi^{(B)}_{A}(Q)\right|_{\dred}\RB &=
   \avbare\,N_c\,\sum_f\,Q_f^2\Lx\frac{4\,\pi}{Q^2\,e^{\gamma_E}}\Rx^\ep
          \left\{ \vphantom{\asbarepi}\right.\\
    & \hskip-50pt
      1 + \asbarepi\Lx\frac{4\,\pi}{Q^2\,e^{\gamma_E}}\Rx^{\ep}
        C_F\,\LB\frac{3}{4} +  \ep\Lx\frac{51}{8}
      - 6\,\zeta_3\Rx   + \ep^2\,\Lx\frac{497}{16} - \frac{15}{4}\zeta_2
           - 15\,\zeta_3 - 9\,\zeta_4\Rx + {\cal O}(\ep^3)\RB\\
    & \hskip-50pt\phantom{1}
       + \aebarepi\Lx\frac{4\,\pi}{Q^2\,e^{\gamma_E}}\Rx^{\ep}
        C_F\,\LB-\ep\,\frac{3}{4} - \ep^2\,\frac{29}{8}
           + {\cal O}(\ep^3)\RB\\
    & \hskip-50pt\phantom{1}
      + \asbarepi^2\,\Lx\frac{4\,\pi}{Q^2\,e^{\gamma_E}}\Rx^{2\,\ep}
      \LB\frac{1}{\ep}\Lx\frac{11}{16}C_F\,C_A
        - \frac{1}{8}C_F\,N_f\Rx - \frac{3}{32}C_F^2
        + \Lx\frac{77}{8} - \frac{33}{4}\zeta_3\Rx\,C_F\,C_A
        - \Lx\frac{7}{4} - \frac{3}{2}\zeta_3\Rx\,C_F\,N_f \right.\\
    &\hskip -30pt +\ep\Lx C_F^2\Lx - \frac{141}{32} - \frac{111}{8}\,\zeta_3
                    + \frac{45}{2}\,\zeta_5\Rx
        + C_F\,C_A\Lx \frac{15301}{192} - \frac{231}{32}\,\zeta_2 - \frac{193}{4}\,\zeta_3
                    - \frac{99}{8}\,\zeta_4 -
                 \frac{15}{4}\,\zeta_5\Rx\right.\\
    &\hskip -20pt\left.\left.
        + C_F\,N_f\Lx - \frac{1355}{96} + \frac{21}{16}\,\zeta_2
                    + \frac{17}{2}\,\zeta_3 +
     \frac{9}{4}\,\zeta_4\Rx\Rx + {\cal O}(\ep^2)\RB\\
    & \hskip-50pt\phantom{1}
      + \aebarepi^2\,\Lx\frac{4\,\pi}{Q^2\,e^{\gamma_E}}\Rx^{2\,\ep}
      \LB\frac{3}{4}C_F^2 - \frac{3}{8}C_F\,C_A
        + \frac{3}{16}C_F\,N_f
        - \ep\Lx\frac{47}{8}C_F^2 - \frac{11}{4}C_F\,C_A
        + \frac{7}{4}C_F\,N_f\Rx + {\cal O}(\ep^2)\RB\\
    &  \hskip-50pt\phantom{1}\left.
      + \asbarepi\aebarepi\,\Lx\frac{4\,\pi}{Q^2\,e^{\gamma_E}}\Rx^{2\,\ep}
      \LB - \frac{9}{8}C_F^2 - \ep\Lx\frac{141}{16}C_F^2
      + \frac{21}{16}C_F\,C_A\Rx + {\cal O}(\ep^2)\RB
         + {\cal O}\Lx\Lx\frac{\asbare}{\pi},\frac{\aebare}{\pi}\Rx^3\Rx\right\}\,,
\label{eqn:impi0Adred}
\end{split}
\end{equation}
and
\begin{equation}
\begin{split}
\Im\LB\left.\Pi^{(B)}_{B}(Q)\right|_{\dred}\RB &=
    \avebare\,N_c\,\sum_f\,Q_f^2\Lx\frac{4\,\pi}{Q^2\,e^{\gamma_E}}\Rx^\ep
          \left\{\ep + 2\,\ep^2 + \Lx4-\frac{3}{2}\zeta_2\Rx\ep^3
         +  {\cal O}(\ep^4)\right.\\
    & \hskip-50pt\phantom{1}
      + \asbarepi\Lx\frac{4\,\pi}{Q^2\,e^{\gamma_E}}\Rx^{\ep}
        C_F\,\LB\frac{3}{2} +  \ep\frac{29}{4} + \ep^2\,\Lx\frac{227}{8} - \frac{15}{2}\zeta_2
           - 6\,\zeta_3\Rx + {\cal O}(\ep^3)\RB\\
    & \hskip-50pt\phantom{1}
       + \aebarepi\Lx\frac{4\,\pi}{Q^2\,e^{\gamma_E}}\Rx^{\ep}
        C_F\,\LB - 1 - 4\,\ep - \ep^2\Lx\frac{27}{2} - 5\,\zeta_2\Rx
           + {\cal O}(\ep^3)\RB\\
    & \hskip-50pt\phantom{1}
      + \asbarepi^2\,\Lx\frac{4\,\pi}{Q^2\,e^{\gamma_E}}\Rx^{2\,\ep}
      \LB\frac{1}{\ep}\Lx\frac{9}{8}C_F^2 + \frac{11}{16}C_F\,C_A
        - \frac{1}{8}C_F\,N_f\Rx
        + \frac{279}{32}C_F^2
        + \frac{199}{32}C_F\,C_A - \frac{17}{16}C_F\,N_f\right. \\
    &\hskip -30pt +\ep\Lx C_F^2\Lx \frac{3139}{64} - \frac{189}{16}\,\zeta_2
                    - \frac{45}{4}\,\zeta_3\Rx
        + C_F\,C_A\Lx \frac{2473}{64} - \frac{231}{32}\,\zeta_2
                    - \frac{75}{8}\,\zeta_3\Rx\right.\\
    &\hskip -20pt\left.\left.
        + C_F\,N_f\Lx - \frac{207}{32} + \frac{21}{16}\,\zeta_2
                    + \frac{3}{2}\,\zeta_3\Rx\Rx + {\cal O}(\ep^2)\RB\\
    &  \hskip-50pt\phantom{1}
      + \asbarepi\aebarepi\,\Lx\frac{4\,\pi}{Q^2\,e^{\gamma_E}}\Rx^{2\,\ep}
      \LB - \frac{1}{\ep}\frac{9}{4}C_F^2 - \frac{129}{8}C_F^2
      - \frac{3}{8}C_F\,C_A\right.\\
    &  \hskip-30pt\phantom{1}\left.
      - \ep\Lx\Lx\frac{671}{8}
      - \frac{189}{8}\zeta_2 - 9\,\zeta_3\Rx\,C_F^2
      + \frac{53}{16}C_F\,C_A\Rx + {\cal O}(\ep^2)\RB\\
    & \hskip-50pt\phantom{1}
      + \aebarepi^2\,\Lx\frac{4\,\pi}{Q^2\,e^{\gamma_E}}\Rx^{2\,\ep}
      \LB\frac{1}{\ep}\Lx C_F^2 - \frac{1}{4}C_F\,C_A + \frac{1}{8}C_F\,N_f\Rx
        + \frac{13}{2}C_F^2 - \frac{3}{2}C_F\,C_A
        + \frac{15}{16}C_F\,N_f\right.\\
    & \hskip-50pt\phantom{1}\left.
        + \ep\Lx\Lx31 - \frac{21}{2}\zeta_2
        - \frac{3}{4}\zeta_3\Rx\,C_F^2 - \Lx\frac{53}{8}
        - \frac{21}{8}\zeta_2 - \frac{3}{8}\zeta_3\Rx\,C_F\,C_A
        + \Lx\frac{157}{32} - \frac{21}{16}\zeta_2\Rx\,C_F\,N_f\Rx
        + {\cal O}(\ep^2)\RB\\
    & \hskip-50pt\phantom{1}\left.
      + {\cal O}\Lx\Lx\frac{\asbare}{\pi},\frac{\aebare}{\pi}\Rx^3\Rx\right\}\,,\\
\label{eqn:impi0Bdred}
\end{split}
\end{equation}
where
$\displaystyle
{\cal O}\Lx\Lx\frac{\asbare}{\pi},\frac{\aebare}{\pi}\Rx^3\Rx
$ denotes terms for which the sum of the powers of
$\displaystyle\Lx\frac{\asbare}{\pi}\Rx$ and
$\displaystyle\Lx\frac{\aebare}{\pi}\Rx$ is at least three.

Upon renormalization according to \eqn{eqn:dredrenorm} and expanding
in terms of $\amsbar$ according to \eqn{eqn:drbartomsbar}, I find that
\begin{equation}
\begin{split}
\Im\LB\left.\Pi_{A}(Q)\right|_{\dred}\RB &=
   \av\,N_c\,\sum_f\,Q_f^2 \left\{
      1 + \adrbarpi\frac{3}{4}C_F
        \LB1+\adrbarpi\betadred{20}\ln\frac{\mu^2}{Q^2}\RB\right.\\
    & \hskip-50pt\phantom{1}\left.
       + \adrbarpi^2\LB-C_F^2\frac{3}{32}
       + C_F\,C_A\Lx\frac{121}{32} - \frac{11}{4}\zeta_3\Rx
       + C_F\,N_f\Lx-\frac{11}{16} + \frac{1}{2}\zeta_3\Rx\RB
       + {\cal O}\Lx\Lx\frac{\asbare}{\pi},\frac{\aebare}{\pi}\Rx^3\Rx\right\}\\
  &\hskip-45pt=\av\,N_c\,\sum_f\,Q_f^2\,
    \left\{  1 + \amsbarpi\, C_F\,\frac{3}{4}\LB1
       + \amsbarpi\,\betabar{0}\,\ln\frac{\mu^2}{Q^2}\RB\right.\\
    & \hskip-45pt\left.
     +\amsbarpi^2\,\LB -C_F^2\,\frac{3}{32}
   + C_F\,C_A\,\Lx\frac{123}{32} - \frac{11}{4}\zeta_3\Rx
   + C_F\,N_f\,\Lx-\frac{11}{16} + \frac{1}{2}\zeta_3\Rx
       \RB  + {\cal O}\Lx\amsbarpi^3\Rx\right\}\,,\\
\Im\LB\left.\Pi_{B}(Q)\right|_{\dred}\RB &={\cal O}(\ep)\,.
\label{eqn:impidred}
\end{split}
\end{equation}

\subsection{Total Decay rate and annihilation cross section in the
  \dred\ scheme}
\label{sec:dreddecann}
As in the \cdr\ scheme, the decay rate and annihilation cross section
are determined from the imaginary part of the forward scattering
amplitude.
\begin{equation}
\Gamma^{\dred}_{V\to\ {\rm hadrons}} = \frac{1}{M_V}\frac{1}{N_{\rm spins}}
    \sum_\lambda \ep^{\mu}(Q,\lambda)\,\Im\LB\left.\Pi_{\mu\nu}(Q)
    \right|_{\dred}\RB \,\ep^{\nu}(Q,\lambda)^{*}\,,
\label{eqn:dreddecay}
\end{equation}
where
\begin{equation}
\frac{1}{N_{\rm spins}}\sum_\lambda \ep^{\mu}(Q,\lambda)\,
    \ep^{\nu}(Q,\lambda)^{*}
   = \frac{1}{3}\Lx-\hat{g}^{\mu\nu} + \frac{Q^\mu\,Q^\nu}{M_V^2}
      - \delta^{\mu\nu}\Rx\,.
\label{eqn:dredspinav}
\end{equation}
The evanescent part of the spin average contracts only with the
$\Pi_{B}(Q)$ term, which has been renormalized to be of order $(\ep)$,
so that the result is
\begin{equation}
\begin{split}
\Gamma^{\dred}_{V\to\ {\rm hadrons}} =&
    \frac{\av\,M_V}{3}\,N_c\,\sum_f\,Q_f^2\,
       \left\{  1 + \amsbarpi\, C_F\,\frac{3}{4}\LB1
       + \amsbarpi\,\betabar{0}\,\ln\frac{\mu^2}{Q^2}\RB\right.\\
    & \hskip-45pt\left.
     +\amsbarpi^2\,\LB -C_F^2\,\frac{3}{32}
   + C_F\,C_A\,\Lx\frac{123}{32}
             - \frac{11}{4}\zeta_3\Rx
   + C_F\,N_f\,\Lx-\frac{11}{16}
         + \frac{1}{2}\,\zeta_3\Rx
       \RB  + {\cal O}\Lx\amsbarpi^3\Rx\right\}\,,
\label{eqn:gamdredverify}
\end{split}
\end{equation}
just like in the \cdr\ calculation.

For the annihilation cross section $\sigma_{e^+\,e^-\to\ {\rm
    hadrons}}$, one attaches fermion bilinears to each end of the
vacuum polarization tensor and averages over the spins.
\begin{equation}
\begin{split}
\sigma^{\dred}_{e^+\,e^-\to\ {\rm hadrons}} &= \frac{2}{Q^2}\frac{e^2}{4}
   \sum_{\lambda\,\lambda^{'}} \frac{\mtrxelm{\overline{v}(p_{e^+},\lambda)}
   {\hat{\gamma}^{\mu}}{u(p_{e^-},\lambda^{'})}}{Q^2}\Im\LB\left.\Pi_{\mu\nu}(Q)
    \right|_{\dred,\,\av\to\alpha}\RB\frac{\mtrxelm{\overline{u}(p_{e^-},\lambda^{'})}
    {\hat{\gamma}^{\nu}}{v(p_{e^+},\lambda)}}{Q^2}\\
   &+ \frac{2}{Q^2}\frac{e_{\ell\,e}^2}{4}
   \sum_{\lambda\,\lambda^{'}} \frac{\mtrxelm{\overline{v}(p_{e^+},\lambda)}
   {\bar{\gamma}^{\mu}}{u(p_{e^-},\lambda^{'})}}{Q^2}\Im\LB\left.\Pi_{\mu\nu}(Q)
    \right|_{\dred,\,\av\to\alpha}\RB\frac{\mtrxelm{\overline{u}(p_{e^-},\lambda^{'})}
    {\bar{\gamma}^{\nu}}{v(p_{e^+},\lambda)}}{Q^2}\,,
\label{eqn:dredannav}
\end{split}
\end{equation}
where $e_{\ell\,e}$ represents the coupling of the evanescent photon to
the electron.  Combining the spinor bilinears into traces,
\begin{equation}
\begin{split}
   \frac{1}{2} &\sum_{\lambda\,\lambda^{'}} \mtrxelm{
    \overline{v}(p_{e^+},\lambda)}{\hat{\gamma}^{\mu}}{u(p_{e^-},\lambda^{'})}
    \mtrxelm{\overline{u}(p_{e^-},\lambda^{'})} {\hat{\gamma}^{\nu}}
    {v(p_{e^+},\lambda)} = \frac{1}{2}\Tr{\slashed{p}_{e^+}\,
    \gamma^{\mu}\slashed{p}_{e^-}\,\gamma^{\nu}}
   = \Lx-Q^2\,\hat{g}^{\mu\,\nu} + Q^\mu\,Q^\nu\Rx\\
   \frac{1}{2} &\sum_{\lambda\,\lambda^{'}} \mtrxelm{
    \overline{v}(p_{e^+},\lambda)}{\bar{\gamma}^{\mu}}{u(p_{e^-},\lambda^{'})}
    \mtrxelm{\overline{u}(p_{e^-},\lambda^{'})} {\bar{\gamma}^{\nu}}
    {v(p_{e^+},\lambda)} = \frac{1}{2}\Tr{\slashed{p}_{e^+}\,
    \bar{\gamma}^{\mu}\slashed{p}_{e^-}\,\bar{\gamma}^{\nu}}
   = \Lx-Q^2\,\delta^{\mu\,\nu}\Rx\,
\end{split}
\end{equation}
The final result is
\begin{equation}
\begin{split}
\sigma^{\dred}_{e^+\,e^-\to\ {\rm hadrons}} =&
    \frac{4\pi\,\alpha^2}{3\,Q^2}\,N_c\,\sum_f\,Q_f^2\,
       \left\{  1 + \amsbarpi\, C_F\,\frac{3}{4}\LB1
       + \amsbarpi\,\betabar{0}\,\ln\frac{\mu^2}{Q^2}\RB\right.\\
    & \hskip-45pt\left.
     +\amsbarpi^2\,\LB -C_F^2\,\frac{3}{32}
   + C_F\,C_A\,\Lx\frac{123}{32}
             - \frac{11}{4}\zeta_3\Rx
   + C_F\,N_f\,\Lx-\frac{11}{16}
         + \frac{1}{2}\,\zeta_3\Rx
       \RB  + {\cal O}\Lx\amsbarpi^3\Rx\right\}\,,
\label{eqn:sigdredverify}
\end{split}
\end{equation}
again in agreement with \eqns{eqn:knownresult}{eqn:knownF}.  As
promised, under the \dred\ scheme renormalization program, evanescent
Green functions are rendered finite by renormalization and contribute
to scattering amplitudes at order $(\ep)$.  Also as promised, the
results are completely equivalent to those of the \cdr\ scheme.

\section{The Four-Dimensional Helicity Scheme}
In the four-dimensional helicity scheme, one defines an enlarged
vector space of dimensionality $D_m = 4-2\,\ep$, in which loop momenta
take values, as in the \cdr\ scheme.  In addition, one defines a still
larger vector space, of dimensionality $D_s=4$, in which internal spin
degrees of freedom take values.  The precise rules for the \fdh\
scheme are given in Ref.~\cite{Bern:2002zk}.  They are:
\begin{enumerate}
\item As in ordinary dimensional regularization, all momentum
  integrals are integrated over $D_m$ dimensional momenta.  Metric
  tensors resulting from tensor integrals are $D_m$ dimensional.
\item All ``observed'' external states are taken to be
  four-dimensional, as are their momenta and polarization vectors.
  This facilitates the use of helicity states for observed particles.
\item All ``unobserved'' or internal states are treated as $D_s$
  dimensional, and the $D_s$ dimensional vector space is taken to be
  larger than the $D_m$ dimensional vector space.  Unobserved states
  include virtual states inside of loops, virtual states inside of
  trees as well as external states that have infrared sensitive
  overlaps with other external states.
\item Both the $D_s$ and $D_m$ dimensional vector spaces are larger
  than the standard four-dimensional space-time, so that contraction
  of four-dimensional objects with $D_m$ or $D_s$ dimensional objects
  yields only four-dimensional components.
\end{enumerate}

To keep track of the many vector spaces and their overlapping domains,
I give the result of the contractions of the various metric tensors
with one another,
\begin{equation}
\begin{split}
g^{\mu\nu}\,g_{\mu\nu} &= D_s\,,\qquad\ 
\hat{g}^{\mu\nu}\,\hat{g}_{\mu\nu} = D_m\,,\qquad
\eta^{\mu\nu}\,\eta_{\mu\nu} = 4\,,\qquad
\delta^{\mu\nu}\,\delta_{\mu\nu} = D_x = D_s - D_m\\
g^{\mu\nu}\hat{g}^{\rho}_{\nu} &= \hat{g}^{\mu\rho}\,,\qquad
g^{\mu\nu}\eta^{\rho}_{\nu} = \eta^{\mu\rho}\,,\qquad
\hat{g}^{\mu\nu}\eta^{\rho}_{\nu} = \eta^{\mu\rho}\,,\\
g^{\mu\nu}\delta^{\rho}_{\nu} &= \delta^{\mu\rho}\,,\qquad
\hat{g}^{\mu\nu}\delta^{\rho}_{\nu} = 0\,,\qquad\quad\ 
\eta^{\mu\nu}\delta^{\rho}_{\nu} = 0\,.\\
\label{eqn::fdhmetrics}
\end{split}
\end{equation}

Like the \hv\ scheme, the \fdh\ scheme treats observed states as
four-dimensional.  In inclusive calculations, however, where there are
infrared overlaps among external states, the external states are taken
to be $D_s$ dimensional in the infrared regions.

As in the \dred\ scheme, spin degrees of freedom take values in a
vector space that is larger than that in which momenta take values.
It would seem, therefore, that the same remarks regarding the Ward
Identity and the conclusion that the $D_x = D_s - D_m$ dimensional
components of the gauge fields and their couplings must be considered
as distinct from the $D_m$ dimensional gauge fields and couplings
would apply.  That is not, however, how the \fdh\ scheme is used.  All
field components in the $D_s$ dimensional space are treated as gauge
fields and no distinction is made between the couplings.  It is
common, however, to define an interpolating scheme, the ``$\delta_R$''
scheme, in which $D_s = 4 - 2\,\ep\,\delta_R$.  The parameter $\delta_R$
interpolates between the \hv\ scheme ($\delta_R=1$) and the \fdh\
scheme ($\delta_R=0$). Using this scheme gives one a handle on the
impact of the evanescent degrees of freedom on the result, but not on
the impact of a distinct evanescent coupling.

It is claimed~\cite{Bern:2002zk} that the essential difference between
the \fdh\ and \dred\ schemes is that in the former $D_m > 4$, while in
the latter $D_m < 4$.  It must be this difference, then, that allows
for the very different handling of the evanescent couplings and
degrees of freedom.  We shall see what impact this choice has in the
calculation and discussion below.

\subsection{Renormalization}
I will not give detailed results for the renormalization parameters of
the \fdh\ scheme.  There is no point in doing so because, as I will
show, the rules of the \fdh\ scheme enumerated in the previous section
are not consistent with a successful renormalization program.  The
first sign that there is a problem with the renormalization program
comes in the computation of the one-loop renormalization constants.
In particular, the gluon vacuum polarization tensor splits into two
independent components, $\Pi_A^{\mu\nu} =
\Pi_A(Q^2)\,\Lx(-Q^2\hat{g}^{\mu\nu} + Q^\mu\,Q^\nu\Rx$ and
$\Pi_B^{\mu\nu} = \Pi_B(Q^2)\,\delta^{\mu\nu}$, both of which are
singular.  This is a clear warning that what the \fdh\ scheme calls
the gluon is in fact two distinct sets of degrees of freedom.  If I
ignore $\Pi_B$ and just renormalize $\Pi_A$, I find the usual result
that
\begin{equation}
\betafdh{0} = \frac{11}{12}C_A - \frac{1}{6}N_f\,.
\end{equation}
Note that I also get this result if I take the spin average (trace) of
the full vacuum polarization tensor.  Because $\Pi_B$ is weighted by a
factor of $2\,\ep$, its contribution to the spin average is not
singular.  Because the leading order term in the quantities being
calculated is of order one, and the \nlo\ term of order $\alpha_s$,
this result for the one-loop $\beta$-function is all that is needed to
compute the renormalized cross section at \nnlo.  Furthermore, the
many \nlo\ results that have been obtained using the \fdh\ scheme have
all renormalized using the above result for $\betafdh{0}$.

When I try to proceed to the two-loop beta function, I find that both
$\Pi_A$ and $\Pi_B$ contribute singular terms to the spin-averaged
vacuum polarization, while if I again ignore $\Pi_B$ and renormalize
$\Pi_A$, I obtain the usual value for $\beta_1$,
\begin{equation}
\betafdh{1} = \frac{17}{24}C_A^2 - \frac{5}{24}C_A\,N_f
          - \frac{1}{8}C_F\,N_f\,.
\end{equation}
This seems to be the choice made in Ref.~\cite{Bern:2002zk} as they
quote only the result for terms proportional to $Q^{\mu}Q^{\nu}$,
which would be part of my $\Pi_A$.  Since the standard lore has been
that $\afdhbar$ and $\adrbar$ coincide, at least through second order
corrections, this seems to be the most reasonable choice.
Furthermore, it means that the conversion to $\amsbar$ will
be~\cite{Kunszt:1993sd,Bern:2002zk}
\begin{equation}
  \afdhbar = \amsbar\LB1 + \amsbarpi\frac{C_A}{12}
  + \ldots\RB
\label{eqn:fdhtomsbar}
\end{equation}
As it turns out, it does not matter what choice one makes as even the
one-loop result for $\betafdh{0}$, which seems safe if only because it
is familiar, leads to the violation of unitarity.

\subsection{Vacuum polarization in the \fdh\ scheme}
Leaving aside the question of renormalization beyond one-loop, I will
proceed with the calculation of the $V$-boson vacuum polarization.  In
performing calculations in the \fdh\ scheme, it becomes apparent that
the results are identical, term-by-term. to the calculation in the
\dred\ scheme, except that the evanescent gluons are identified as
gluons and the coupling $\alpha_e$ is set to $\alpha_s$.  Therefore I find
that
\begin{equation}
  \Im\LB\left.\Pi^{(B)}_{\mu\nu}(Q)\right|_{\fdh}\RB =
    \frac{-Q^2\,\hat{g}_{\mu\nu} + Q_{\mu}Q_{\nu}}{3}\,
       \Im\LB\left.\Pi^{(B)}_{A}(Q)\right|_{\fdh}\RB
   - Q^2\,\frac{\delta_{\mu\nu}}{2\,\ep}\,\Im\LB\left.\Pi^{(B)}_{B}(Q)\right|_{\fdh}\RB\,,
\end{equation}
where
\begin{equation}
\begin{split}
\Im\LB\left.\Pi^{(B)}_{A}(Q)\right|_{\fdh}\RB &=
   \avbare\,N_c\,\sum_f\,Q_f^2\Lx\frac{4\,\pi}{Q^2\,e^{\gamma_E}}\Rx^\ep
          \left\{ \vphantom{\asbarepi}\right.\\
    & \hskip-50pt
      1 + \asbarepi\Lx\frac{4\,\pi}{Q^2\,e^{\gamma_E}}\Rx^{\ep}
        C_F\,\LB\frac{3}{4} +  \ep\Lx\frac{45}{8}
      - 6\,\zeta_3\Rx   + \ep^2\,\Lx\frac{439}{16} - \frac{15}{4}\zeta_2
           - 15\,\zeta_3 - 9\,\zeta_4\Rx + {\cal O}(\ep^3)\RB\\
    & \hskip-50pt\phantom{1}
      + \asbarepi^2\,\Lx\frac{4\,\pi}{Q^2\,e^{\gamma_E}}\Rx^{2\,\ep}
      \LB\frac{1}{\ep}\Lx\frac{11}{16}C_F\,C_A
        - \frac{1}{8}C_F\,N_f\Rx - \frac{15}{32}C_F^2
        + \Lx\frac{37}{4} - \frac{33}{4}\zeta_3\Rx\,C_F\,C_A
        - \Lx\frac{25}{16} - \frac{3}{2}\zeta_3\Rx\,C_F\,N_f \right.\\
    &\hskip -30pt +\ep\Lx C_F^2\Lx - \frac{235}{32} - \frac{111}{8}\,\zeta_3
                    + \frac{45}{2}\,\zeta_5\Rx
        + C_F\,C_A\Lx \frac{14521}{192} - \frac{231}{32}\,\zeta_2 - \frac{193}{4}\,\zeta_3
                    - \frac{99}{8}\,\zeta_4 -
                 \frac{15}{4}\,\zeta_5\Rx\right.\\
    &\hskip -20pt\left.\left.\left.
        + C_F\,N_f\Lx - \frac{1187}{96} + \frac{21}{16}\,\zeta_2
                    + \frac{17}{2}\,\zeta_3 +
     \frac{9}{4}\,\zeta_4\Rx\Rx + {\cal O}(\ep^2)\RB
         + {\cal O}\Lx\asbarepi^3\Rx\right\}\,,
\label{eqn:impi0Afdh}
\end{split}
\end{equation}
and
\begin{equation}
\begin{split}
\Im\LB\left.\Pi^{(B)}_{B}(Q)\right|_{\fdh}\RB &=
    \avbare\,N_c\,\sum_f\,Q_f^2\Lx\frac{4\,\pi}{Q^2\,e^{\gamma_E}}\Rx^\ep
          \left\{ \vphantom{\asbarepi}\right.\\
    & \hskip-50pt \ep+
      \asbarepi\Lx\frac{4\,\pi}{Q^2\,e^{\gamma_E}}\Rx^{\ep}
        C_F\,\LB\frac{1}{2} +  \ep\frac{13}{4} + \ep^2\,\Lx\frac{119}{8} - \frac{5}{2}\zeta_2
           - 6\,\zeta_3\Rx + {\cal O}(\ep^3)\RB\\
    & \hskip-50pt\phantom{1}
      + \asbarepi^2\,\Lx\frac{4\,\pi}{Q^2\,e^{\gamma_E}}\Rx^{2\,\ep}
      \LB\frac{1}{\ep}\Lx-\frac{1}{8}C_F^2 + \frac{7}{16}C_F\,C_A\Rx
        - \frac{29}{32}C_F^2
        + \frac{139}{32}C_F\,C_A - \frac{1}{8}C_F\,N_f\right. \\
    &\hskip -30pt +\ep\Lx C_F^2\Lx -\frac{245}{64} + \frac{21}{16}\,\zeta_2
                    - 3\,\zeta_3\Rx
        + C_F\,C_A\Lx \frac{1837}{64} - \frac{147}{32}\,\zeta_2
                    - 9\,\zeta_3\Rx\right.\\
    &\hskip -20pt\left.\left.\left.
        + C_F\,N_f\Lx - \frac{25}{16} + \frac{3}{2}\,\zeta_3\Rx\Rx
         + {\cal O}(\ep^2)\RB
      + {\cal O}\Lx\asbarepi^3\Rx \right\}\,.\\
\label{eqn:impi0Bfdh}
\end{split}
\end{equation}
Upon renormalizing such that 
\begin{equation}
\asbarepi\to\afdhbarpi\Lx\frac{4\,\pi}{Q^2\,e^{\gamma_E}}\Rx^{-\ep}
  \Lx1 - \frac{\betafdh{0}}{\ep}\afdhbarpi\Rx
\,,\qquad\qquad
\avbare\to\av\Lx\frac{4\,\pi}{Q^2\,e^{\gamma_E}}\Rx^{-\ep}\,,
\end{equation}
I find that
\begin{equation}
\begin{split}
\Im\LB\left.\Pi_{A}(Q)\right|_{\fdh}\RB &=
   \av\,N_c\,\sum_f\,Q_f^2 \left\{
      1 + \afdhbarpi\frac{3}{4}C_F
        \LB1+\afdhbarpi\betafdh{0}\ln\frac{\mu^2}{Q^2}\RB\right.\\
    & \hskip-50pt\phantom{1}\left.
       + \afdhbarpi^2\LB-C_F^2\frac{15}{32}
       + C_F\,C_A\Lx\frac{131}{32} - \frac{11}{4}\zeta_3\Rx
       + C_F\,N_f\Lx-\frac{5}{8} + \frac{1}{2}\zeta_3\Rx\RB
       + {\cal O}\Lx\afdhbarpi^3\Rx\right\}\\
  &\hskip-45pt=\av\,N_c\,\sum_f\,Q_f^2\,
    \left\{  1 + \amsbarpi\, C_F\,\frac{3}{4}\LB1
       + \amsbarpi\,\betabar{0}\,\ln\frac{\mu^2}{Q^2}\RB\right.\\
    & \hskip-45pt\left.
     +\amsbarpi^2\,\LB -C_F^2\,\frac{15}{32}
   + C_F\,C_A\,\Lx\frac{133}{32} - \frac{11}{4}\zeta_3\Rx
   + C_F\,N_f\,\Lx-\frac{5}{8} + \frac{1}{2}\zeta_3\Rx
       \RB  + {\cal O}\Lx\amsbarpi^3\Rx\right\}\,,\\
\Im\LB\left.\Pi_{B}(Q)\right|_{\fdh}\RB &=
   \av\,N_c\,\sum_f\,Q_f^2 \left\{
      \afdhbarpi\frac{1}{2}C_F
        \LB1+\afdhbarpi\betafdh{0}\ln\frac{\mu^2}{Q^2}\RB\right.\\
    & \hskip-50pt\phantom{1}
       + \afdhbarpi^2\LB\frac{1}{\ep}\Lx
           - C_F^2\frac{1}{8} - C_F\,C_A\frac{1}{48}
           + C_F\,N_f\frac{1}{12}\Rx\Lx1 + 3\ep\ln\frac{\mu^2}{Q^2}\Rx\right.\\
    &\left.\left.
       - C_F^2\frac{29}{32}
       + C_F\,C_A\frac{131}{96}
       + C_F\,N_f\frac{5}{12}\RB
       + {\cal O}\Lx\afdhbarpi^3\Rx\right\}\\
  &\hskip-45pt=
   \av\,N_c\,\sum_f\,Q_f^2 \left\{
      \amsbarpi\frac{1}{2}C_F
        \LB1+\amsbarpi\betafdh{0}\ln\frac{\mu^2}{Q^2}\RB\right.\\
    & \hskip-50pt\phantom{1}
       + \amsbarpi^2\LB\frac{1}{\ep}\Lx
           - C_F^2\frac{1}{8} - C_F\,C_A\frac{1}{48}
           + C_F\,N_f\frac{1}{12}\Rx\Lx1 + 3\ep\ln\frac{\mu^2}{Q^2}\Rx\right.\\
    &\left.\left.
       - C_F^2\frac{29}{32}
       + C_F\,C_A\frac{45}{32}
       + C_F\,N_f\frac{5}{12}\RB
       + {\cal O}\Lx\amsbarpi^3\Rx\right\}\,.
\label{eqn:impifdh}
\end{split}
\end{equation}

\subsection{Total Decay rate and annihilation cross section in the
  \fdh\ scheme}
\label{sec:fdhdecann}
The results of the vacuum polarization calculation look to be
disastrous as $\Pi_B$ is singular at order $\alpha_s^2$.  However, the
rules of the \fdh\ scheme, enumerated above, specify that external
states are taken to be four-dimensional.  This means that the spin
average of the vector polarizations is
\begin{equation}
\frac{1}{N_{\rm spins}}\sum_\lambda \ep^{\mu}(Q,\lambda)\,
    \ep^{\nu}(Q,\lambda)^{*}
   = \frac{1}{3}\Lx-\eta^{\mu\nu} + \frac{Q^\mu\,Q^\nu}{M_V^2}\Rx\,,
\label{eqn:fdhspinav}
\end{equation}
which annihilates $\left.\Pi_B^{\mu\nu}\right|_\fdh$.  For the
annihilation rate, the rules are a bit ambiguous, as they could be
read to mean that the lepton spinors are four-dimensional but the
vertex ($\gamma^\mu$) connecting them to the loop part of the
amplitude is $D_s$ dimensional.  This would bring
$\left.\Pi_B^{\mu\nu}\right|_\fdh$ into the calculation and lead to a
singular result at order $\alpha_s^2$.  However, Rule $4$ could also
be read to mean that the vertex sandwiched between four-dimensional
states is also reduced to being four-dimensional.

Assuming this interpretation, I find that
\begin{equation}
\begin{split}
\Gamma^{\fdh}_{V\to\ {\rm hadrons}} =&
    \frac{\av\,M_V}{3}\,N_c\,\sum_f\,Q_f^2\,
       \left\{  1 + \amsbarpi\, C_F\,\frac{3}{4}\LB1
       + \amsbarpi\,\betabar{0}\,\ln\frac{\mu^2}{Q^2}\RB\right.\\
    & \hskip-45pt\left.
     +\amsbarpi^2\,\LB -C_F^2\,\frac{15}{32}
   + C_F\,C_A\,\Lx\frac{133}{32}
             - \frac{11}{4}\zeta_3\Rx
   + C_F\,N_f\,\Lx-\frac{5}{8}
         + \frac{1}{2}\,\zeta_3\Rx
       \RB  + {\cal O}\Lx\amsbarpi^3\Rx\right\}\,,
\label{eqn:gamfdhverify}
\end{split}
\end{equation}
and
\begin{equation}
\begin{split}
\sigma^{\fdh}_{e^+\,e^-\to\ {\rm hadrons}} =&
    \frac{4\pi\,\alpha^2}{3\,Q^2}\,N_c\,\sum_f\,Q_f^2\,
       \left\{  1 + \amsbarpi\, C_F\,\frac{3}{4}\LB1
       + \amsbarpi\,\betabar{0}\,\ln\frac{\mu^2}{Q^2}\RB\right.\\
    & \hskip-45pt\left.
     +\amsbarpi^2\,\LB -C_F^2\,\frac{15}{32}
   + C_F\,C_A\,\Lx\frac{133}{32}
             - \frac{11}{4}\zeta_3\Rx
   + C_F\,N_f\,\Lx-\frac{5}{8}
         + \frac{1}{2}\,\zeta_3\Rx
       \RB  + {\cal O}\Lx\amsbarpi^3\Rx\right\}\,.
\label{eqn:sigfdhverify}
\end{split}
\end{equation}
The results agree with one another, are correct through \nlo\ and are
finite through \nnlo.  Unfortunately, the \nnlo\ terms are not
correct!  Because the discrepancy is finite, there remains the
possibility that the conversion from $\afdhbar$ to $\amsbar$ given in
\eqn{eqn:fdhtomsbar} is incorrect, although this would contradict
previous results~\cite{Kunszt:1993sd,Bern:2002zk}.  If this were the
case, then one would expect that the \nnnlo\ result would also be
finite but incorrect.  If, instead, the finite discrepancy at \nnlo\
is the result of a failure of the renormalization program, the \nnnlo\
result should be singular.

\section{Partial results at \nnnlo}
\begin{figure}[h]
\includegraphics[width=4.cm]{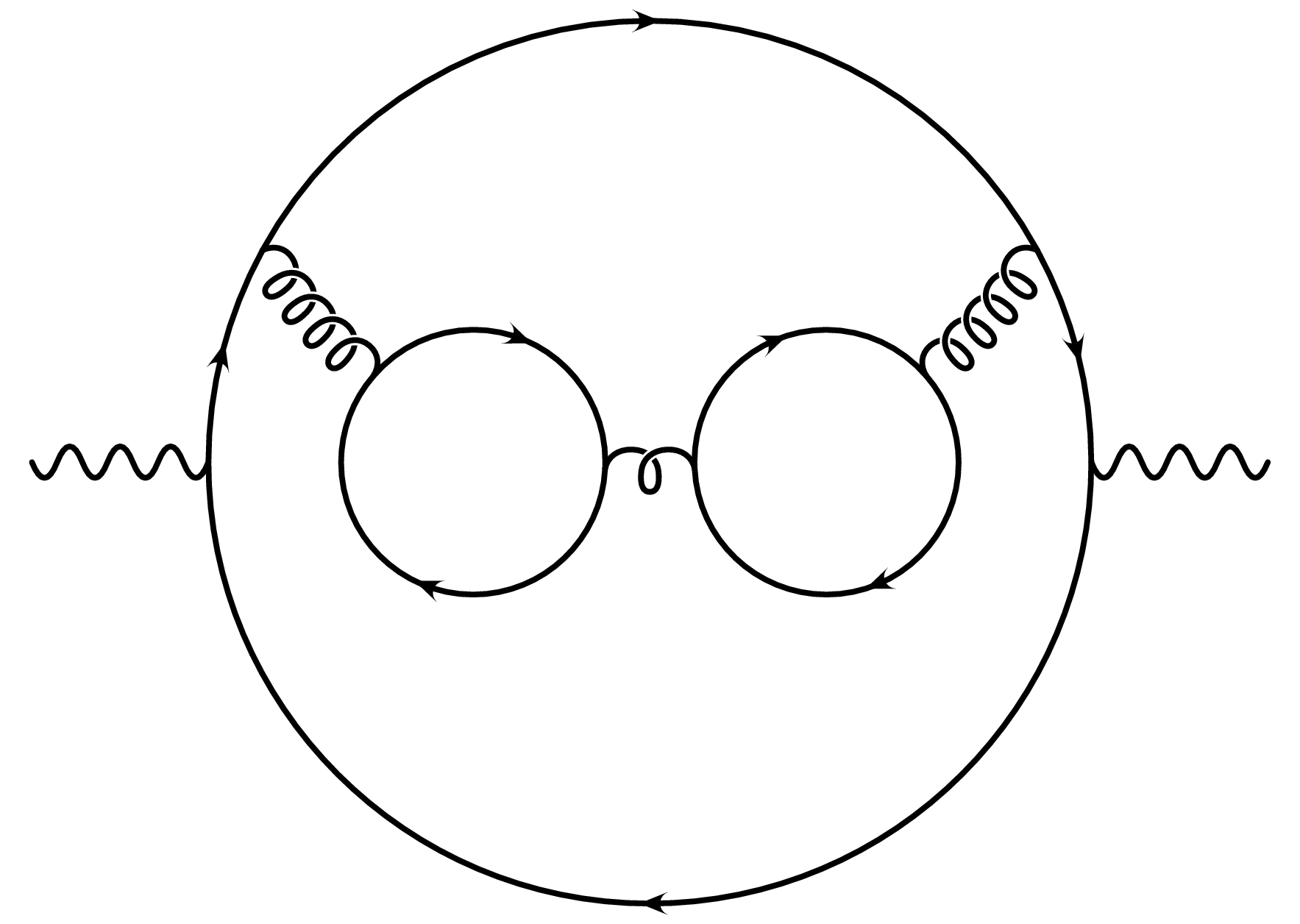}
\includegraphics[width=4.cm]{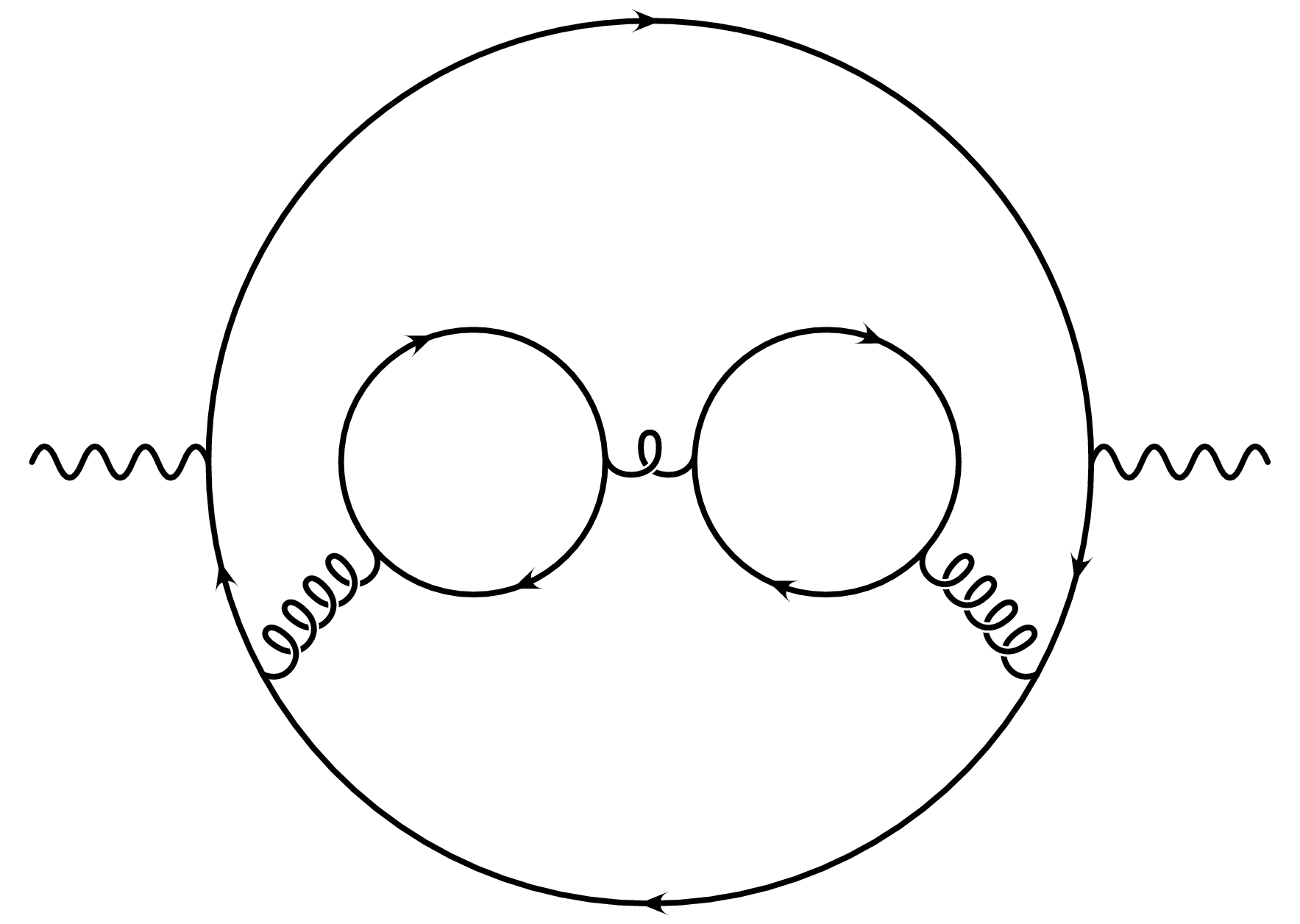}
\includegraphics[width=4.cm]{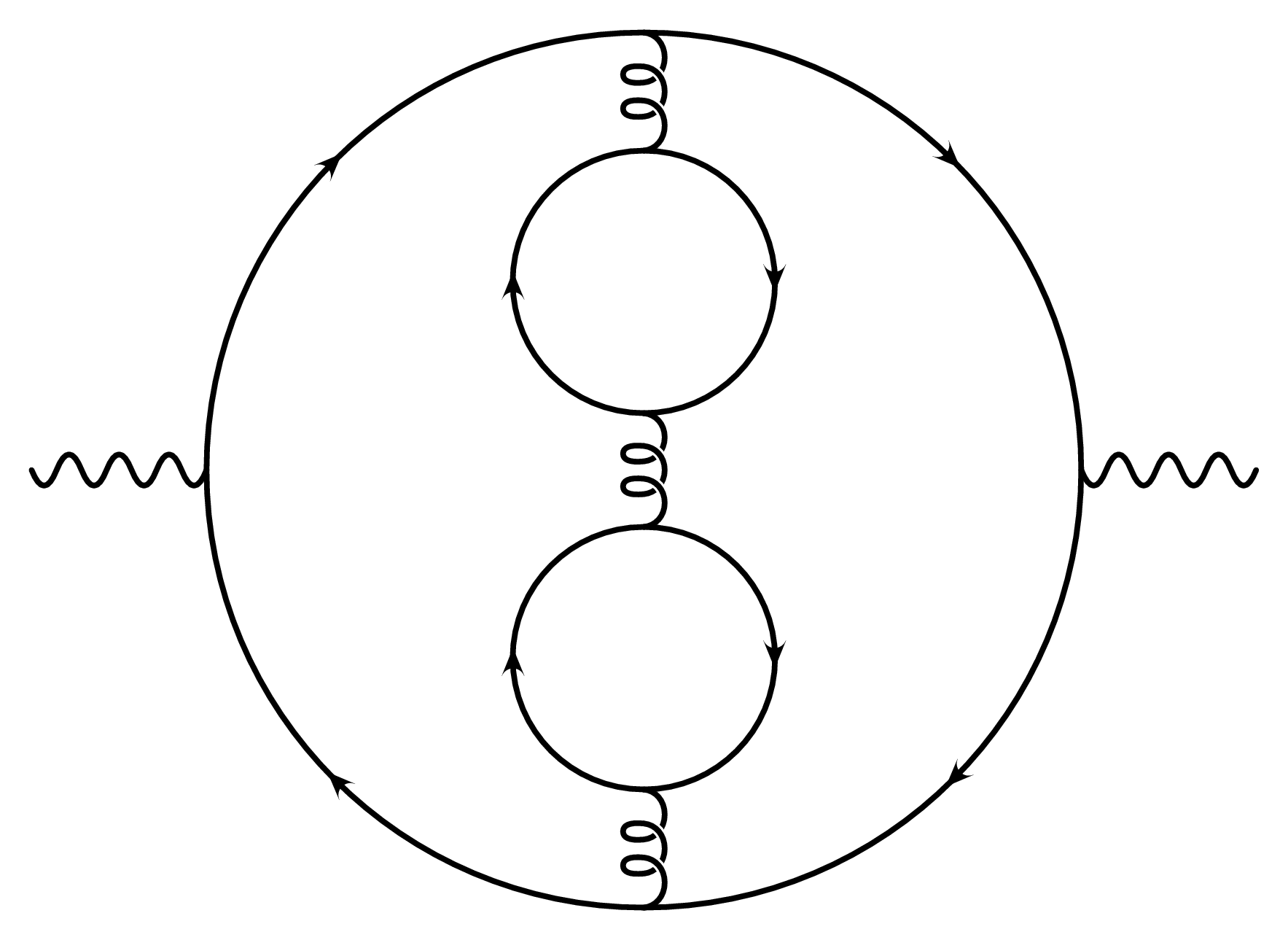}
\caption{Four-loop diagrams that contribute to the $N_f^2$ term at
 \nnnlo.
\label{fig:fourloop}}
\end{figure}

Although first computed some time ago, the vacuum polarization at four
loops~\cite{Gorishnii:1988bc,Gorishnii:1990vf} remains a formidable
calculation.  It is only necessary, however, to look at a small part
of the calculation: the terms proportional to the square of the number
of fermion flavors, $N_f^2$.  This is fortunate for a couple of reasons:
1) there are only three four-loop diagrams to be computed, see
\fig{fig:fourloop}, (plus three more in the \dred\ scheme, where the
gluons are replaced by evanescent gluons); and 2) the contributions
from renormalization in the \cdr\ and \fdh\ schemes come only from the
leading term in the \qcd\ $\beta$-function ($\beta_{0}$ and
$\beta_{0}^2$).  Thus, my result will not depend on how the higher
order terms of the $\beta$-function are chosen in the \fdh\ scheme.

\subsection{The \cdr\ scheme}
In the \cdr\ scheme, there are only three four-loop diagrams that need
to be calculated.  The first two are simply iterated-bubble diagrams
and are essentially trivial.  The third is slightly nontrivial, so I
again use my QGRAF-FORM-REDUZE suite of programs to address the
problem.  All of the four-loop master integrals can be found in
Ref.~\cite{Baikov:2010hf}. I find the result of the four-loop
calculation to be
\begin{equation}
\begin{split}
\Im\LB\left.\Pi^{(B)}_{\mu\nu}(Q)\right|_{\cdr}\RB_{\alpha_s^3\,N_f^2}
    &=\frac{ -Q^2\,g_{\mu\nu} + Q_{\mu}Q_{\nu}}{3}
   \avbare\,N_c\,\sum_f\,Q_f^2\Lx\frac{4\,\pi}{Q^2\,e^{\gamma_E}}
   \Rx^{4\,\ep}\\
  &\qquad\times \asbarepi^3\,C_F\,N_f^2\LB\frac{1}{48\,\ep^2}
   + \frac{1}{\ep}\Lx\frac{121}{288}
   - \frac{1}{3}\zeta_3\Rx + \frac{2777}{576} - \frac{3}{8}\zeta_2
   - \frac{19}{6}\zeta_3 - \frac{1}{2}\zeta_4\RB
\label{eqn:fourloopcdrbare}
\end{split}
\end{equation}
Renormalizing, I find
\begin{equation}
\begin{split}
\Im\LB\left.\Pi_{\mu\nu}(Q)\right|_{\cdr}\RB_{\alpha_s^3\,N_f^2}
    &=\frac{ -Q^2\,g_{\mu\nu} + Q_{\mu}Q_{\nu}}{3}
   \av\,N_c\,\sum_f\,Q_f^2\amsbarpi^3\,C_F\,N_f^2\\
  &\times\LB
  \frac{151}{216} - \frac{1}{24}\zeta_2
   - \frac{19}{36}\zeta_3 + \Lx\frac{11}{48} - \frac{1}{6}\zeta_3\Rx
    \ln\Lx\frac{\mu^2}{Q^2}\Rx + \frac{1}{48}\ln^2\Lx\frac{\mu^2}{Q^2}\Rx\RB
\label{eqn:fourloopcdr}
\end{split}
\end{equation}
Using this term to compute the $\alpha_s^3\,N_f^2$ contribution to the
decay rate and annihilation cross section as in
Eqs.~(\ref{eqn:cdrdecay},\ref{eqn:cdrannav}), I find the result
expected from \eqns{eqn:knownresult}{eqn:knownF}.

\subsection{The \dred\ scheme}
In the \dred\ scheme, there are three extra four-loop diagrams to
compute, obtained by replacing gluon propagators with evanescent gluon
propagators.  I find
\begin{equation}
\begin{split}
\Im\LB\left.\Pi^{(B)}_{A}(Q)\right|_{\dred}\RB_{\alpha_s^3\,N_f^2} &=
   \avbare\,N_c\,\sum_f\,Q_f^2\Lx\frac{4\,\pi}{Q^2\,e^{\gamma_E}}\Rx^{4\,\ep}
          C_F\,N_f^2\,\left\{ \vphantom{\asbarepi}\right.\\
    &\phantom{+}
       \asbarepi^3\,\LB\frac{1}{48\,\ep^2} + \frac{1}{\ep}\Lx
       \frac{13}{32} - \frac{1}{3}\zeta_3\Rx
      + \frac{7847}{1728} - \frac{3}{8}\zeta_2
      - \frac{53}{18}\zeta_3 - \frac{1}{2}\zeta_4\RB\\
    &\left.+ \aebarepi^3\,\LB-\frac{1}{\ep}\frac{3}{64} - \frac{83}{128}\RB
        \right\}\\
\Im\LB\left.\Pi^{(B)}_{B}(Q)\right|_{\dred}\RB_{\alpha_s^3\,N_f^2} &=
    \avebare\,N_c\,\sum_f\,Q_f^2\Lx\frac{4\,\pi}{Q^2\,e^{\gamma_E}}\Rx^{4\,\ep}
          C_F\,N_f^2\,\left\{ \vphantom{\asbarepi}\right.\\
    & \phantom{+}\asbarepi^3\,\LB\frac{1}{72\,\ep^2}
          + \frac{1}{\ep}\frac{73}{432} + \frac{3595}{2592} 
          - \frac{1}{4}\zeta_2 - \frac{1}{3}\zeta_3\RB\\
    &\left.+ \aebarepi^3\,\LB-\frac{1}{48\,\ep^2}
          - \frac{1}{\ep}\frac{11}{48} - \frac{155}{96}
          + \frac{3}{8}\zeta_2\RB \right\}
\label{eqn:impi40dred}
\end{split}
\end{equation}
Upon renormalizing according to \eqn{eqn:dredrenorm} and converting
the coupling to $\amsbar$, I obtain
\begin{equation}
\begin{split}
\Im\LB\left.\Pi_{A}(Q)\right|_{\dred}\RB_{\alpha_s^3\,N_f^2} &\\
  &\hskip -55pt =
   \av\,N_c\,\sum_f\,Q_f^2\,C_F\,N_f^2\amsbarpi^3\,\LB
  \frac{151}{216} - \frac{1}{24}\zeta_2
   - \frac{19}{36}\zeta_3 + \Lx\frac{11}{48} - \frac{1}{6}\zeta_3\Rx
    \ln\Lx\frac{\mu^2}{Q^2}\Rx
    + \frac{1}{48}\ln^2\Lx\frac{\mu^2}{Q^2}\Rx\RB\,,\\
\Im\LB\left.\Pi_{B}(Q)\right|_{\dred}\RB_{\alpha_s^3\,N_f^2} &={\cal O}(\ep)\,.
\label{eqn:impi4dred}
\end{split}
\end{equation}
As for the \cdr\ scheme, this leads to the expected result for the
decay rate and annihilation cross section.

\subsection{The \fdh\ scheme}
In the \fdh\ scheme, however, I find that
\begin{equation}
\begin{split}
\Im\LB\left.\Pi^{(B)}_{A}(Q)\right|_{\fdh}\RB_{\alpha_s^3\,N_f^2} &=
   \avbare\,N_c\,\sum_f\,Q_f^2\Lx\frac{4\,\pi}{Q^2\,e^{\gamma_E}}\Rx^{4\,\ep}
          C_F\,N_f^2\\
    &\times
       \asbarepi^3\,\LB\frac{1}{48\,\ep^2} + \frac{1}{\ep}\Lx
       \frac{23}{64} - \frac{1}{3}\zeta_3\Rx
      + \frac{13453}{3456} - \frac{3}{8}\zeta_2
      - \frac{53}{18}\zeta_3 - \frac{1}{2}\zeta_4\RB\,,\\
\Im\LB\left.\Pi^{(B)}_{B}(Q)\right|_{\fdh}\RB_{\alpha_s^3\,N_f^2} &=
    \avbare\,N_c\,\sum_f\,Q_f^2\Lx\frac{4\,\pi}{Q^2\,e^{\gamma_E}}\Rx^{4\,\ep}
          C_F\,N_f^2\\
    & \times\asbarepi^3\,\LB-\frac{1}{144\,\ep^2}
          - \frac{1}{\ep}\frac{13}{216} - \frac{295}{1296} 
          + \frac{1}{8}\zeta_2 - \frac{1}{3}\zeta_3\RB\,.
\label{eqn:impi40fdh}
\end{split}
\end{equation}
I renormalize according to
\begin{equation}
\asbare =\Lx\frac{\mu^2\,e^{\gamma_E}}{4\,\pi}\Rx^\ep\,
    \afdhbar\,\LB1 - \afdhbarpi\frac{\betafdh{0}}{\ep}
        + \afdhbarpi^2\Lx\frac{\betafdh{0}^2}{\ep^2}
        - \frac{1}{2}\frac{\betafdh{1}}{\ep}\Rx\RB\,,
\end{equation}
keeping only terms proportional to $\afdhbar^3\,N_f^2$.  Such terms
can only come from the $\betafdh{0}$ and $\betafdh{0}^2$ terms, so any
uncertainty about $\betafdh{1}$ has no effect here.
The renormalized result is
\begin{equation}
\begin{split}
\Im\LB\left.\Pi_{A}(Q)\right|_{\fdh}\RB_{\alpha_s^3\,N_f^2} &\\
  &\hskip -75pt =
   \av\,N_c\,\sum_f\,Q_f^2\,C_F\,N_f^2\afdhbarpi^3\,\LB
  -\frac{1}{192\,\ep} + \frac{1843}{3456} - \frac{1}{24}\zeta_2
   - \frac{19}{36}\zeta_3 + \Lx\frac{3}{16} - \frac{1}{6}\zeta_3\Rx
    \ln\Lx\frac{\mu^2}{Q^2}\Rx
    + \frac{1}{48}\ln^2\Lx\frac{\mu^2}{Q^2}\Rx\RB\,,\\
\Im\LB\left.\Pi_{B}(Q)\right|_{\fdh}\RB_{\alpha_s^3\,N_f^2} &\\
  &\hskip -75pt =
   \av\,N_c\,\sum_f\,Q_f^2\,C_F\,N_f^2\afdhbarpi^3\,\LB
    \frac{1}{144\,\ep^2} -\frac{5}{432\,\ep} - \frac{869}{2592}
    + \frac{1}{18}\zeta_2 - \frac{5}{27}\ln\Lx\frac{\mu^2}{Q^2}\Rx
    - \frac{1}{36}\ln^2\Lx\frac{\mu^2}{Q^2}\Rx\RB\,.
\label{eqn:impi4fdh}
\end{split}
\end{equation}
The demand that external states be four-dimensional removes the
$\Pi_B$ term, but there is also a pole in $\Pi_A$ and no finite
renormalization to put the result in terms of $\amsbar$ can remove it.
I must therefore conclude that the \fdh\ scheme is not consistent with
unitarity.

\section{Discussion}
In this paper, I have performed a high-order calculation in each of
three regularization schemes: the conventional dimensional
regularization (\cdr) scheme; the dimensional reduction (\dred)
scheme; and the four-dimensional helicity (\fdh) scheme.  Of these,
the \cdr\ scheme is by far the most widely used, and was, in fact,
used to compute the original results that I use as my test basis.  The
\fdh\ scheme has primarily been used to produce one-loop helicity
amplitudes, although it has been used in a few cases in two-loop
calculations and also as a supersymmetric regulator.  The primary
purpose of this paper was to put the \fdh\ scheme to a stringent test
and determine its reliability in a high-order calculation.  The \dred\
scheme is primarily used as a supersymmetric regulator and is quite
cumbersome for nonsupersymmetric calculations.  It is, however,
closely related to the \fdh\ scheme and has been
demonstrated~\cite{Jack:1993ws,
Jack:1994bn,Harlander:2006rj,Harlander:2006xq} to be equivalent to the
\cdr\ scheme through four loops.  A close comparison of the details of
the calculations in the \fdh\ and \dred\ schemes helps to identify
where and when things go wrong with the former.

In the cases of the \cdr\ and \dred\ schemes, I have reproduced the
known result for the hadronic decay width of a massive vector boson
(or equivalently, the $e^+e^-$ annihilation rate to hadrons) through
\nnlo, and a few terms at \nnnlo.  This represents computing the
\qcd\ corrections to the vacuum polarization of the photon ($V$
boson) through three loops, with partial results at four loops.  In
addition, I have reproduced the renormalization parameters of \qcd\
($\beta$-function(s), mass anomalous dimension) through three-loop
order.  This establishes that I have theoretical control over all of
the needed calculations through three-loop order.  In order to obtain
the partial \nnnlo\ result in the \dred\ scheme, I also needed the
three-loop \qcd\ corrections to the $\beta$-function of the evanescent
photon ($V$ boson).

The calculation of the $V$ boson decay rate provides another instance
of the equivalence the \cdr\ and \dred\ schemes at the four-loop
level~\cite{Harlander:2006xq}.  The ability to obtain the correct
result using the \dred\ scheme required a delicate balance of the many
extra couplings and their renormalization effects upon one another.
Indeed, given the complexity needed to make the \dred\ scheme work, it
seems that there should be little surprise that the \fdh\ scheme, with
its greater simplicity, should fail.  

Perhaps, it is worth considering how it is that the \fdh\ scheme has
been used successfully in so many calculations.  Its most common use
has been in the construction of one-loop scattering amplitudes via
unitarity cuts, using four-dimensional helicity amplitudes as the
primary building blocks.  Thus, it is natural that it restricts
observed (external) states to be four-dimensional.  Because the \fdh\
scheme defines that $D_s > D_m > 4$, this restriction excludes
evanescent fields from appearing as external states.  This is very
important because, as one can see from comparing
Eqs.~(\ref{eqn:impidred}) and (\ref{eqn:impi4fdh}), terms involving
external evanescent states are the most dangerous.  Even though it
does not renormalize evanescent states and couplings properly the
\fdh\ is able to get the nonevanescent part of the vacuum
polarization tensor correct at \nlo, while the evanescent part is
ready to contribute a finite error at \nlo.  Because the \dred\ scheme
defines $4 > D_m$, the evanescent states are {\it parts} of the
classical four-dimensional states.  It would not seem natural to
exclude them from appearing as external states.  Instead, they are
handled through the renormalization program so that their effects are
removed from physical scattering amplitudes.  In the \fdh\ scheme, the
evanescent states are instead {\it additions} to the four-dimensional
states (as are the extra degrees of freedom that come from
regularizing momentum integrals) and there is no barrier to excluding
them as observed states.

In an \fdh\ scheme calculation, a tree-level term is strictly
four-dimensional and is free from evanescent contributions.
(Depending on interpretation, this may be a stronger condition than is
given in the rules of Ref~\cite{Bern:2002zk}, but it is the actual
condition imposed if one defines the tree-level amplitude as being a
four-dimensional helicity amplitude.)  Because evanescent terms are
absent at tree-level, they cannot generate ultraviolet poles at one
loop.  Even if one were to renormalize them properly, as in the \dred\
scheme, there would be nowhere to make the counter-term insertion!  In
fact, the one-loop contributions are not even finite, as the counting
over the number of states ($2\ep$) makes the result of order $\ep$.
This is clearly illustrated in \eqn{eqn:impi0Adred}.  Neither
$\alpha_s$, nor $\alpha_e$ appear at \lo.  Therefore, the
contributions at \nlo\ are finite for $\alpha_s$ and of order $\ep$
(because of the counting over the number of states) for $\alpha_e$.
In more complicated \qcd\ calculations, $\alpha_s$ will appear at \lo\
and will therefore contribute an ultraviolet pole at one-loop, which
will be removed by renormalization.  $\alpha_e$, however, will still
make its first appearance at \nlo\ and that contribution will be of
order $\ep$.  Thus, one can expect that the \fdh\ scheme, used as
above, should be reliable for computing \nlo\ corrections through
finite order ($\ep^0$).  The error from improperly identifying evanescent
quantities should be of order $\ep$.  At \nnlo\ and beyond however,
the failure to properly identify and renormalize the evanescent
parameters leads to incorrect results and the violation of unitarity.

So, as suggested~\cite{Bern:2002zk}, one of the \fdh\ scheme's most
important assets is that it defines $D_s > D_m > 4$.  This feature is
also the scheme's undoing, though not of necessity.  Because the
effects of external evanescent states can be removed (or indeed never
seen) by imposing a four-dimensionality restriction, and because the
effects of internal evanescent states therefore contribute at order
$\ep$ at one loop, it appears that one can simply ignore the
distinction between gauge and evanescent terms.  In contrast, because
the \dred\ scheme must deal with external evanescent terms from the
beginning, its advocates were forced to develop a successful
renormalization program~\cite{Jack:1993ws, Jack:1994bn}.  Extensive
testing~\cite{Jack:1993ws,Jack:1994bn,Harlander:2006rj,
Harlander:2006xq} has shown that this program works to at least the
fourth order and that it handles the effects of both internal and
external evanescent contributions.  As I remarked earlier,
calculations in the \dred\ and \fdh\ schemes are term-by-term
identical, except for the identification of the couplings and
propagating states.  Thus, one could make the \fdh\ scheme a unitary
regularization scheme for nonsupersymmetric calculations by
recognizing the distinction between gauge and evanescent terms and
adopting the \dred\ scheme's renormalization program.  This would, of
course, do away with any notion of the \fdh\ scheme being simple, but
it would at least be correct.  The \fdh\ scheme would still be
distinguished from the \dred\ scheme by the fact that $D_s > D_m > 4$,
which facilitates helicity amplitude calculations and, in chiral
theories, improves its situation with regard to $\gamma_5$ and the
Levi-Civita tensor~\cite{Siegel:1980qs,Stockinger:2005gx}.
Furthermore, with a valid renormalization program, the requirement of
four-dimensional observed states could be made optional.  This would
lead to two linked, slightly different, schemes, just like the \hv\
and \cdr\ schemes.  This suggestion has already been made by Signer
and St\"ockinger~\cite{Signer:2008va} who in fact define their version
of the \dred\ scheme to have precisely the $D_s > D_m > 4$ hierarchy
of the \fdh\ scheme.

Thus, in conclusion, the \cdr\ and \dred\ schemes are correct and
equivalent ways of performing \qcd\ calculations through \nnnlo.  The
\fdh\ scheme, however, has been shown to be incorrect and to violate
unitarity beyond \nlo\ when applied to nonsupersymmetric theories.  It
must therefore be viewed as a shortcut for performing \nlo\
calculations and should only be used for such calculations with great
caution.

\vskip20pt

\paragraph*{Acknowledgments:}
This research was supported by the U.S.~Department of Energy under
Contract No.~DE-AC02-98CH10886.

\appendix
\section{Renormalization parameters for the \cdr\ scheme}
\label{sec:cdrrenorm}
To three-loop order, I find the coefficients of the $\beta$-function
to be
\begin{equation}
\begin{split}
    \betabar{0} &= \frac{11}{12}C_A - \frac{1}{6}N_f\,,\qquad\qquad
    \betabar{1} = \frac{17}{24}C_A^2 - \frac{5}{24}C_A\,N_f
          - \frac{1}{8}C_F\,N_f\,,\\
    \betabar{2} &= \frac{2857}{3456}C_A^3
        - \frac{1415}{3456}C_A^2\,N_f - \frac{205}{1152}C_A\,C_F\,N_f
        + \frac{1}{64}C_F^2\,N_f + \frac{79}{3456}C_A\,N_f^2
        + \frac{11}{576}C_F\,N_f^2\,,
\label{eqn:cdrbeta}
\end{split}
\end{equation}
while the coefficients of the mass anomalous dimension are
\begin{equation}
\begin{split}
    \gammabar{0} &= \frac{3}{4}C_F\,,\hskip80pt
    \gammabar{1} = \frac{3}{32}C_F^2 + \frac{97}{96}C_F\,C_A
                - \frac{5}{48}C_F\,N_f\,,\\
    \gammabar{2} &= \frac{129}{128}C_F^3 - \frac{129}{256}C_F^2\,C_A
        + \frac{11413}{6912}C_F\,C_A^2
        - \Lx\frac{23}{64} - \frac{3}{8}\zeta_3\Rx\,C_F^2\,N_f
         - \Lx\frac{139}{864} + \frac{3}{8}\zeta_3\Rx\,C_F\,C_A\,N_f
         - \frac{35}{1728}C_F\,N_f^2\,,
\label{eqn:cdrgamma}
\end{split}
\end{equation}
in agreement with known results~\cite{Tarasov:1980au,Larin:1993tp,
Chetyrkin:1997dh,Vermaseren:1997fq}.

\section{Renormalization parameters for the \dred\ scheme}
\label{sec:dredrenorm}
The coefficients of the \qcd\ $\beta$-function, $\betadred{}(\adrbar)$
through three loops are:
\begin{equation}
\begin{split}
    \betadred{20} &= \frac{11}{12}C_A - \frac{1}{6}N_f\,,\hskip60pt
    \betadred{30} = \frac{17}{24}C_A^2 - \frac{5}{24}C_A\,N_f
      - \frac{1}{8}C_F\,N_f\,,\\
    \betadred{40} &= \frac{3115}{3456}C_A^3
      - \frac{1439}{3456}C_A^2\,N_f - \frac{193}{1152}C_A\,C_F\,N_f
      + \frac{1}{64}C_F^2\,N_f + \frac{79}{3456}C_A\,N_f^2
      + \frac{11}{576}C_F\,N_f^2\,,\\
   \betadred{31} & = - \frac{1}{16}C_F\,N_f\Lx\frac{3}{2}C_F\Rx\,,\qquad\quad
   \betadred{22} = - \frac{1}{16}C_F\,N_f\Lx \frac{1}{2}C_A - C_F
      - \frac{1}{4}N_f\Rx\,,
\label{eqn:dredbeta}
\end{split}
\end{equation}
where the notation is that
\begin{equation}
\betadred{}(\adrbar) = -\ep\frac{\adrbar}{\pi} - \sum_{i,j,k,l,m}\,\betadred{ijklm}
    \,\adrbarpi^{i}\,\aedrbarpi^{j}\,\etadrbarpi{1}^{k}
    \,\etadrbarpi{2}^{l}\,\etadrbarpi{3}^{m}\,.
\end{equation}
The last three indices of $\betadred{ijklm}$ are omitted when
they are all equal to $0$.

The $\beta$-function of evanescent \qcd\ coupling,
$\betaedred{}{}(\aedrbar)$ is
\begin{equation}
\begin{split}
    \betaedred{0}{2} &= \frac{1}{2}C_A - C_F - \frac{1}{4}N_f\,,\qquad
    \betaedred{1}{1} = \frac{3}{2}C_F\,,\\[5pt]
    \betaedred{0}{3} &= \frac{3}{8}C_A^2 - \frac{5}{4}C_A\,C_F
             + C_F^2 - \frac{3}{16}C_A\,N_f
             + \frac{3}{8}C_F\,N_f\,,\qquad
    \betaedred{1}{2} =  - \frac{3}{8}C_A^2 + \frac{5}{2}C_A\,C_F
              - \frac{11}{4}C_F^2 - \frac{5}{16}C_F\,N_f\,,\\
    \betaedred{2}{1} &=  - \frac{7}{64}C_A^2
              +\frac{55}{48}C_A\,C_F + \frac{3}{16}C_F^2
              + \frac{1}{16}C_A\,N_f - \frac{5}{24}C_F\,N_f\\[5pt]
    \betaedred{0}{2100} \hskip-13pt&\hskip13pt= -\frac{9}{8}\qquad
    \betaedred{0}{2010} = \frac{5}{4}\qquad
    \betaedred{0}{2001} = \frac{3}{4}\\
    \betaedred{0}{1200} \hskip-13pt&\hskip13pt= \frac{27}{64}\qquad
    \betaedred{0}{1020} = -\frac{15}{4}\qquad
    \betaedred{0}{1002} = \frac{21}{32}\qquad
    \betaedred{0}{1101} = -\frac{9}{16}\\
    \betaedred{0}{4} &=
              - \Lx\frac{7}{4} + \frac{9}{4}\zeta_3\Rx\,C_F^3
              + \Lx\frac{17}{8} + \frac{15}{2}\zeta_3\Rx\,C_F^2\,C_A
              - \Lx\frac{3}{4} + \frac{69}{16}\zeta_3\Rx\,C_F\,C_A^2
              + \Lx\frac{1}{16} + \frac{9}{16}\zeta_3\Rx\,C_A^3\\
         &
              + \Lx\frac{13}{32} - \frac{33}{16}\zeta_3\Rx\,C_F^2\,N_f
              + \Lx\frac{1}{32} + \frac{51}{32}\zeta_3\Rx\,C_F\,C_A\,N_f
              - \Lx\frac{21}{128} + \frac{9}{32}\zeta_3\Rx\,C_A^2\,N_f
              - \Lx\frac{1}{128}C_F - \frac{7}{256}C_A\Rx\,N_f^2\\
    \betaedred{1}{3} &=
                \Lx\frac{13}{2} - 3\,\zeta_3\Rx\,C_F^3
              - \Lx10 - 6\,\zeta_3\Rx\,C_F^2\,C_A
              + \Lx\frac{133}{32} - \frac{15}{4}\zeta_3\Rx\,C_F\,C_A^2
              - \Lx\frac{25}{64} - \frac{3}{4}\zeta_3\Rx\,C_A^3\\
         &
              + \Lx\frac{13}{16} - \frac{3}{4}\zeta_3\Rx\,C_F^2\,N_f
              - \frac{9}{8}\Lx1-\zeta_3\Rx\,C_F\,C_A\,N_f
              + \Lx\frac{7}{32} - \frac{3}{8}\zeta_3\Rx\,C_A^2\,N_f
              + \frac{3}{64}\,C_A\,N_f^2\\
    \betaedred{2}{2} &= 
              - \Lx\frac{139}{64} - \frac{27}{4}\zeta_3\Rx\,C_F^3
              - \Lx\frac{793}{128} + 18\,\zeta_3\Rx\,C_F^2\,C_A
              + \Lx\frac{1587}{256} + \frac{207}{16}\zeta_3\Rx\,C_F\,C_A^2
              - \Lx\frac{427}{512} + \frac{45}{16}\zeta_3\Rx\,C_A^3\\
         &
              - \Lx\frac{569}{256} - \frac{99}{16}\zeta_3\Rx\,C_F^2\,N_f
              + \Lx\frac{31}{16} - \frac{171}{32}\zeta_3\Rx\,C_F\,C_A\,N_f
              - \Lx\frac{871}{1024} - \frac{45}{32}\zeta_3\Rx\,C_A^2\,N_f
              + \Lx\frac{1}{16}C_F - \frac{1}{256}C_A\Rx\,N_f^2\\
    \betaedred{3}{1} &= 
                \frac{129}{64}C_F^3 - \frac{457}{128}C_F^2\,C_A
              + \frac{11875}{3456}C_F\,C_A^2
              - \frac{3073}{4608}C_A^3\\
         &
              - \Lx\frac{23}{32} - \frac{3}{4}\zeta_3\Rx\,C_F^2\,N_f
              - \Lx\frac{157}{1728} + \frac{3}{4}\zeta_3\Rx\,C_F\,C_A\,N_f
              + \frac{463}{2304}C_A^2\,N_f
              - \Lx\frac{35}{864}C_F + \frac{5}{576}C_A\Rx\,N_f^2\\
    \betaedred{0}{3100} \hskip-13pt&\hskip13pt =
                   - \frac{9}{64} + \frac{243}{128}N_f\qquad
    \betaedred{0}{3010} = \frac{5}{8} - \frac{45}{64}N_f\qquad
    \betaedred{0}{3001} = \frac{3}{32} - \frac{81}{64}N_f\\
    \betaedred{1}{2100} \hskip-13pt&\hskip13pt = -\frac{219}{16}\qquad
    \betaedred{1}{2010} = \frac{145}{48}\qquad
    \betaedred{1}{2001} = \frac{73}{8}\\
    \betaedred{2}{1100} \hskip-13pt&\hskip13pt = -\frac{1125}{1024}\qquad
    \betaedred{2}{1010} = \frac{105}{128}\qquad
    \betaedred{2}{1001} = \frac{615}{512}\\
    \betaedred{0}{2200} \hskip-13pt&\hskip13pt =
                  \frac{1413}{512} - \frac{729}{1024}N_f\qquad
    \betaedred{0}{2020} = - \frac{115}{32} + \frac{135}{64}N_f\qquad
    \betaedred{0}{2002} = - \frac{161}{256} - \frac{567}{512}N_f\\
    \betaedred{0}{2110} \hskip-13pt&\hskip13pt =
                 \frac{75}{8}\qquad
    \betaedred{0}{2101} = -\frac{471}{128} + \frac{243}{256}N_f\qquad
    \betaedred{0}{2011} = -\frac{85}{8}\\
    \betaedred{0}{1300} \hskip-13pt&\hskip13pt = -\frac{1701}{1024}\qquad
    \betaedred{0}{1210} = -\frac{405}{128}\qquad
    \betaedred{0}{1201} = \frac{1701}{512}\\
    \betaedred{0}{1120} \hskip-13pt&\hskip13pt = \frac{135}{32}\qquad
    \betaedred{0}{1111} = \frac{135}{16}\qquad
    \betaedred{0}{1102} = -\frac{81}{128}\\
    \betaedred{0}{1021} \hskip-13pt&\hskip13pt =
                   - \frac{315}{32}\qquad
    \betaedred{0}{1012} = - \frac{315}{32}\qquad
    \betaedred{0}{1003} = \frac{63}{128}\qquad
\label{eqn:dredbetae}
\end{split}
\end{equation}

The mass anomalous dimension in the \dred\ scheme is
\begin{equation}
\begin{split}
\gammadred{10} &= \frac{3}{4}C_F \\[5pt]
\gammadred{20} &= \frac{3}{32}C_F^2 + \frac{91}{96}C_A\,C_F
     - \frac{5}{48}C_F\,N_f\qquad
\gammadred{11} = - \frac{3}{8}C_F^2 \qquad
\gammadred{02} = \frac{1}{4}C_F^2 - \frac{1}{8}C_A\,C_F
     + \frac{1}{16}C_F\,N_f\\[5pt]
\gammadred{30} &= \frac{129}{128}C_F^3 - \frac{133}{256}C_F^2\,C_A
     + \frac{10255}{6912}C_F\,C_A^2
     - \Lx\frac{23}{64} - \frac{3}{8}\zeta_3\Rx\,C_F^2\,N_f
     - \Lx\frac{281}{1728} + \frac{3}{8}\zeta_3\Rx\,C_A\,C_F\,N_f
     - \frac{35}{1728}C_F\,N_f^2\\
\gammadred{21} &= - \frac{27}{64}C_F^3 - \frac{21}{32}C_F^2\,C_A
     - \frac{15}{256}C_F\,C_A^2 + \frac{9}{64}C_F^2\,N_f\\
\gammadred{12} &= \frac{9}{8}C_F^3 - \frac{21}{32}C_F^2\,C_A
     + \frac{3}{64}C_F\,C_A^2 + \frac{3}{128}C_F\,C_A\,N_f
     + \frac{3}{16}C_F^2\,N_f\\
\gammadred{03} &= - \frac{3}{8}C_F^3 + \frac{3}{8}C_F^2\,C_A
     - \frac{3}{32}C_F\,C_A^2 + \frac{1}{16}C_F\,C_A\,N_f
     - \frac{5}{32}C_F^2\,N_f - \frac{1}{128}C_F\,N_f^2\\[5pt]
\gammadred{02100} \hskip-10pt&\hskip10pt= \frac{3}{8}\qquad
\gammadred{02010} = -\frac{5}{12}\qquad
\gammadred{02001} = -\frac{1}{4}\qquad\\
\gammadred{01200} \hskip-10pt&\hskip10pt= -\frac{9}{64}\qquad
\gammadred{01101} = \frac{3}{16}\qquad
\gammadred{01020} = \frac{5}{4}\qquad
\gammadred{01002} = - \frac{7}{32}
\label{eqn:dredgamma}
\end{split}
\end{equation}
The above results for $\betadred{}$, $\betaedred{}{}$ and $\gammadred{}$
all agree with the results of Refs.~\cite{Harlander:2006rj,
Harlander:2006xq}

The \qcd\ contributions to the $\beta$-function of the evanescent part
of a non-\qcd\ gauge coupling is a new result.  I find
\begin{equation}
\begin{split}
\betavedred{1}{0}& = \frac{3}{2}C_F\qquad \betavedred{0}{1} = - C_F\\[5pt]
\betavedred{2}{0}& = \frac{3}{16}C_F^2 + \frac{91}{48}C_F\,C_A
     - \frac{5}{24}C_F\,N_f \qquad
\betavedred{1}{1} = - \frac{11}{4}C_F^2 - \frac{3}{4}C_F\,C_A\qquad
\betavedred{0}{2} = C_F^2 + \frac{3}{8}C_F\,N_f\\[5pt]
\betavedred{3}{0} & = \frac{129}{64}C_F^3
    - \frac{133}{128}C_F^2\,C_A
    - \Lx\frac{23}{32} - \frac{3}{4}\zeta_3\Rx\,C_F^2\,N_f
    + \frac{10255}{3456}C_F\,C_A^2
    - \Lx\frac{281}{864} + \frac{3}{4}\zeta_3\Rx\,C_F\,C_A\,N_f
    - \frac{35}{864}C_F\,N_f^2\\
\betavedred{2}{1} & = - \Lx\frac{139}{64} - \frac{27}{4}\zeta_3\Rx\,C_F^3
    - \Lx\frac{331}{64} + \frac{81}{8}\zeta_3\Rx\,C_F^2\,C_A
    + \frac{11}{16}C_F^2\,N_f
    - \Lx\frac{195}{256} - \frac{27}{8}\zeta_3\Rx\,C_F\,C_A^2
    + \frac{5}{64}C_F\,C_A\,N_f\\
\betavedred{1}{2} & = \Lx\frac{13}{2} - 3\,\zeta_3\Rx\,C_F^3
    - \Lx\frac{7}{8} - \frac{9}{2}\zeta_3\Rx\,C_F^2\,C_A
    + \Lx\frac{63}{64} - \frac{3}{4}\zeta_3\Rx\,C_F^2\,N_f
    + \Lx\frac{7}{16} - \frac{3}{2}\zeta_3\Rx\,C_F\,C_A^2\\
         &
    - \Lx\frac{3}{64} - \frac{3}{4}\zeta_3\Rx\,C_F\,C_A\,N_f\\[5pt]
\betavedred{0}{3} & = - \Lx\frac{7}{4} + \frac{9}{4}\zeta_3\Rx\,C_F^3
    + \Lx\frac{1}{8} + \frac{27}{8}\zeta_3\Rx\,C_F^2\,C_A
    - \frac{27}{32}C_F^2\,N_f
    + \Lx\frac{1}{16} - \frac{9}{8}\zeta_3\Rx\,C_F\,C_A^2
    + \frac{3}{64}C_F\,C_A\,N_f
    + \frac{3}{64}C_F\,N_f^2\\
\betavedred{0}{2100} \hskip -13pt&\hskip13pt= \frac{3}{8}\qquad
\betavedred{0}{2010} = -\frac{25}{6}\qquad
\betavedred{0}{2001} = -\frac{1}{4}\\
\betavedred{0}{1200} \hskip -13pt&\hskip13pt=-\frac{63}{64}\qquad
\betavedred{0}{1101} = \frac{21}{16}\qquad
\betavedred{0}{1020} = \frac{65}{4}\qquad
\betavedred{0}{1002} = -\frac{49}{32}
\label{eqn:dredbetave}
\end{split}
\end{equation}


\begin{thebibliography}{33}
\expandafter\ifx\csname natexlab\endcsname\relax\def\natexlab#1{#1}\fi
\expandafter\ifx\csname bibnamefont\endcsname\relax
  \def\bibnamefont#1{#1}\fi
\expandafter\ifx\csname bibfnamefont\endcsname\relax
  \def\bibfnamefont#1{#1}\fi
\expandafter\ifx\csname citenamefont\endcsname\relax
  \def\citenamefont#1{#1}\fi
\expandafter\ifx\csname url\endcsname\relax
  \def\url#1{\texttt{#1}}\fi
\expandafter\ifx\csname urlprefix\endcsname\relax\def\urlprefix{URL }\fi
\providecommand{\bibinfo}[2]{#2}
\providecommand{\eprint}[2][]{\url{#2}}

\bibitem[{\citenamefont{'t~Hooft and Veltman}(1972)}]{'tHooft:1972fi}
\bibinfo{author}{\bibfnamefont{G.}~\bibnamefont{'t~Hooft}} \bibnamefont{and}
  \bibinfo{author}{\bibfnamefont{M.~J.~G.} \bibnamefont{Veltman}},
  \bibinfo{journal}{Nucl. Phys.} \textbf{\bibinfo{volume}{B44}},
  \bibinfo{pages}{189} (\bibinfo{year}{1972}).

\bibitem[{\citenamefont{Collins}(1984)}]{Collins:Renorm}
\bibinfo{author}{\bibfnamefont{J.}~\bibnamefont{Collins}},
  \emph{\bibinfo{title}{Renormalization}} (\bibinfo{publisher}{Cambridge
  University Press}, \bibinfo{address}{Cambridge, England},
  \bibinfo{year}{1984}).

\bibitem[{\citenamefont{Siegel}(1979)}]{Siegel:1979wq}
\bibinfo{author}{\bibfnamefont{W.}~\bibnamefont{Siegel}},
  \bibinfo{journal}{Phys. Lett.} \textbf{\bibinfo{volume}{B84}},
  \bibinfo{pages}{193} (\bibinfo{year}{1979}).

\bibitem[{\citenamefont{Bern and Kosower}(1992)}]{Bern:1992aq}
\bibinfo{author}{\bibfnamefont{Z.}~\bibnamefont{Bern}} \bibnamefont{and}
  \bibinfo{author}{\bibfnamefont{D.~A.} \bibnamefont{Kosower}},
  \bibinfo{journal}{Nucl. Phys.} \textbf{\bibinfo{volume}{B379}},
  \bibinfo{pages}{451} (\bibinfo{year}{1992}).

\bibitem[{\citenamefont{Bern et~al.}(2002)\citenamefont{Bern, De~Freitas,
  Dixon, and Wong}}]{Bern:2002zk}
\bibinfo{author}{\bibfnamefont{Z.}~\bibnamefont{Bern}},
  \bibinfo{author}{\bibfnamefont{A.}~\bibnamefont{De~Freitas}},
  \bibinfo{author}{\bibfnamefont{L.~J.} \bibnamefont{Dixon}}, \bibnamefont{and}
  \bibinfo{author}{\bibfnamefont{H.~L.} \bibnamefont{Wong}},
  \bibinfo{journal}{Phys. Rev.} \textbf{\bibinfo{volume}{D66}},
  \bibinfo{pages}{085002} (\bibinfo{year}{2002}), \eprint{hep-ph/0202271}.

\bibitem[{\citenamefont{van Damme and 't~Hooft}(1985)}]{vanDamme:1984ig}
\bibinfo{author}{\bibfnamefont{R.}~\bibnamefont{van Damme}} \bibnamefont{and}
  \bibinfo{author}{\bibfnamefont{G.}~\bibnamefont{'t~Hooft}},
  \bibinfo{journal}{Phys. Lett.} \textbf{\bibinfo{volume}{B150}},
  \bibinfo{pages}{133} (\bibinfo{year}{1985}).

\bibitem[{\citenamefont{Capper et~al.}(1980)\citenamefont{Capper, Jones, and
  van Nieuwenhuizen}}]{Capper:1980ns}
\bibinfo{author}{\bibfnamefont{D.~M.} \bibnamefont{Capper}},
  \bibinfo{author}{\bibfnamefont{D.~R.~T.} \bibnamefont{Jones}},
  \bibnamefont{and} \bibinfo{author}{\bibfnamefont{P.}~\bibnamefont{van
  Nieuwenhuizen}}, \bibinfo{journal}{Nucl. Phys.}
  \textbf{\bibinfo{volume}{B167}}, \bibinfo{pages}{479} (\bibinfo{year}{1980}).

\bibitem[{\citenamefont{Jack et~al.}(1994{\natexlab{a}})\citenamefont{Jack,
  Jones, and Roberts}}]{Jack:1993ws}
\bibinfo{author}{\bibfnamefont{I.}~\bibnamefont{Jack}},
  \bibinfo{author}{\bibfnamefont{D.~R.~T.} \bibnamefont{Jones}},
  \bibnamefont{and} \bibinfo{author}{\bibfnamefont{K.~L.}
  \bibnamefont{Roberts}}, \bibinfo{journal}{Z. Phys.}
  \textbf{\bibinfo{volume}{C62}}, \bibinfo{pages}{161}
  (\bibinfo{year}{1994}{\natexlab{a}}), \eprint{hep-ph/9310301}.

\bibitem[{\citenamefont{Jack et~al.}(1994{\natexlab{b}})\citenamefont{Jack,
  Jones, and Roberts}}]{Jack:1994bn}
\bibinfo{author}{\bibfnamefont{I.}~\bibnamefont{Jack}},
  \bibinfo{author}{\bibfnamefont{D.~R.~T.} \bibnamefont{Jones}},
  \bibnamefont{and} \bibinfo{author}{\bibfnamefont{K.~L.}
  \bibnamefont{Roberts}}, \bibinfo{journal}{Z. Phys.}
  \textbf{\bibinfo{volume}{C63}}, \bibinfo{pages}{151}
  (\bibinfo{year}{1994}{\natexlab{b}}), \eprint{hep-ph/9401349}.

\bibitem[{\citenamefont{Chetyrkin et~al.}(1979)\citenamefont{Chetyrkin, Kataev,
  and Tkachov}}]{Chetyrkin:1979bj}
\bibinfo{author}{\bibfnamefont{K.~G.} \bibnamefont{Chetyrkin}},
  \bibinfo{author}{\bibfnamefont{A.~L.} \bibnamefont{Kataev}},
  \bibnamefont{and} \bibinfo{author}{\bibfnamefont{F.~V.}
  \bibnamefont{Tkachov}}, \bibinfo{journal}{Phys. Lett.}
  \textbf{\bibinfo{volume}{B85}}, \bibinfo{pages}{277} (\bibinfo{year}{1979}).

\bibitem[{\citenamefont{Dine and Sapirstein}(1979)}]{Dine:1979qh}
\bibinfo{author}{\bibfnamefont{M.}~\bibnamefont{Dine}} \bibnamefont{and}
  \bibinfo{author}{\bibfnamefont{J.~R.} \bibnamefont{Sapirstein}},
  \bibinfo{journal}{Phys. Rev. Lett.} \textbf{\bibinfo{volume}{43}},
  \bibinfo{pages}{668} (\bibinfo{year}{1979}).

\bibitem[{\citenamefont{Celmaster and Gonsalves}(1980)}]{Celmaster:1980ji}
\bibinfo{author}{\bibfnamefont{W.}~\bibnamefont{Celmaster}} \bibnamefont{and}
  \bibinfo{author}{\bibfnamefont{R.~J.} \bibnamefont{Gonsalves}},
  \bibinfo{journal}{Phys. Rev.} \textbf{\bibinfo{volume}{D21}},
  \bibinfo{pages}{3112} (\bibinfo{year}{1980}).

\bibitem[{\citenamefont{Gorishnii et~al.}(1988)\citenamefont{Gorishnii, Kataev,
  and Larin}}]{Gorishnii:1988bc}
\bibinfo{author}{\bibfnamefont{S.~G.} \bibnamefont{Gorishnii}},
  \bibinfo{author}{\bibfnamefont{A.~L.} \bibnamefont{Kataev}},
  \bibnamefont{and} \bibinfo{author}{\bibfnamefont{S.~A.} \bibnamefont{Larin}},
  \bibinfo{journal}{Phys. Lett.} \textbf{\bibinfo{volume}{B212}},
  \bibinfo{pages}{238} (\bibinfo{year}{1988}).

\bibitem[{\citenamefont{Gorishnii et~al.}(1991)\citenamefont{Gorishnii, Kataev,
  and Larin}}]{Gorishnii:1990vf}
\bibinfo{author}{\bibfnamefont{S.~G.} \bibnamefont{Gorishnii}},
  \bibinfo{author}{\bibfnamefont{A.~L.} \bibnamefont{Kataev}},
  \bibnamefont{and} \bibinfo{author}{\bibfnamefont{S.~A.} \bibnamefont{Larin}},
  \bibinfo{journal}{Phys. Lett.} \textbf{\bibinfo{volume}{B259}},
  \bibinfo{pages}{144} (\bibinfo{year}{1991}).

\bibitem[{\citenamefont{Nogueira}(1993)}]{Nogueira:1993ex}
\bibinfo{author}{\bibfnamefont{P.}~\bibnamefont{Nogueira}},
  \bibinfo{journal}{J. Comput. Phys.} \textbf{\bibinfo{volume}{105}},
  \bibinfo{pages}{279} (\bibinfo{year}{1993}).

\bibitem[{\citenamefont{Vermaseren}(2000)}]{Vermaseren:2000nd}
\bibinfo{author}{\bibfnamefont{J.~A.~M.} \bibnamefont{Vermaseren}}
  (\bibinfo{year}{2000}), \bibinfo{note}{{Report} {No.} {NIKHEF}-00-0032},
  \eprint[http://arXiv.org/abs]{math-ph/0010025}.

\bibitem[{\citenamefont{Studerus}(2010)}]{Studerus:2009ye}
\bibinfo{author}{\bibfnamefont{C.}~\bibnamefont{Studerus}},
  \bibinfo{journal}{Comput. Phys. Commun.} \textbf{\bibinfo{volume}{181}},
  \bibinfo{pages}{1293} (\bibinfo{year}{2010}), \eprint{0912.2546}.

\bibitem[{\citenamefont{Davydychev et~al.}(1998)\citenamefont{Davydychev,
  Osland, and Tarasov}}]{Davydychev:1997vh}
\bibinfo{author}{\bibfnamefont{A.~I.} \bibnamefont{Davydychev}},
  \bibinfo{author}{\bibfnamefont{P.}~\bibnamefont{Osland}}, \bibnamefont{and}
  \bibinfo{author}{\bibfnamefont{O.}~\bibnamefont{Tarasov}},
  \bibinfo{journal}{Phys.Rev.} \textbf{\bibinfo{volume}{D58}},
  \bibinfo{pages}{036007} (\bibinfo{year}{1998}), \eprint{hep-ph/9801380}.

\bibitem[{\citenamefont{Chetyrkin et~al.}(1980)\citenamefont{Chetyrkin, Kataev,
  and Tkachov}}]{Chetyrkin:1980pr}
\bibinfo{author}{\bibfnamefont{K.~G.} \bibnamefont{Chetyrkin}},
  \bibinfo{author}{\bibfnamefont{A.~L.} \bibnamefont{Kataev}},
  \bibnamefont{and} \bibinfo{author}{\bibfnamefont{F.~V.}
  \bibnamefont{Tkachov}}, \bibinfo{journal}{Nucl. Phys.}
  \textbf{\bibinfo{volume}{B174}}, \bibinfo{pages}{345} (\bibinfo{year}{1980}).

\bibitem[{\citenamefont{Kazakov}(1984)}]{Kazakov:1983ns}
\bibinfo{author}{\bibfnamefont{D.~I.} \bibnamefont{Kazakov}},
  \bibinfo{journal}{Theor. Math. Phys.} \textbf{\bibinfo{volume}{58}},
  \bibinfo{pages}{223} (\bibinfo{year}{1984}).

\bibitem[{\citenamefont{Gorishnii et~al.}(1989)\citenamefont{Gorishnii, Larin,
  Surguladze, and Tkachov}}]{Gorishnii:1989gt}
\bibinfo{author}{\bibfnamefont{S.~G.} \bibnamefont{Gorishnii}},
  \bibinfo{author}{\bibfnamefont{S.~A.} \bibnamefont{Larin}},
  \bibinfo{author}{\bibfnamefont{L.~R.} \bibnamefont{Surguladze}},
  \bibnamefont{and} \bibinfo{author}{\bibfnamefont{F.~V.}
  \bibnamefont{Tkachov}}, \bibinfo{journal}{Comput. Phys. Commun.}
  \textbf{\bibinfo{volume}{55}}, \bibinfo{pages}{381} (\bibinfo{year}{1989}).

\bibitem[{\citenamefont{Larin et~al.}(1991)\citenamefont{Larin, Tkachov, and
  Vermaseren}}]{Larin:1991fz}
\bibinfo{author}{\bibfnamefont{S.~A.} \bibnamefont{Larin}},
  \bibinfo{author}{\bibfnamefont{F.~V.} \bibnamefont{Tkachov}},
  \bibnamefont{and} \bibinfo{author}{\bibfnamefont{J.~A.~M.}
  \bibnamefont{Vermaseren}} (\bibinfo{year}{1991}), \bibinfo{note}{{Report}
  {No.} {NIKHEF}-H-91-18}.

\bibitem[{\citenamefont{Harlander
  et~al.}(2006{\natexlab{a}})\citenamefont{Harlander, Jones, Kant, Mihaila, and
  Steinhauser}}]{Harlander:2006xq}
\bibinfo{author}{\bibfnamefont{R.~V.} \bibnamefont{Harlander}},
  \bibinfo{author}{\bibfnamefont{D.~R.~T.} \bibnamefont{Jones}},
  \bibinfo{author}{\bibfnamefont{P.}~\bibnamefont{Kant}},
  \bibinfo{author}{\bibfnamefont{L.}~\bibnamefont{Mihaila}}, \bibnamefont{and}
  \bibinfo{author}{\bibfnamefont{M.}~\bibnamefont{Steinhauser}},
  \bibinfo{journal}{JHEP} \textbf{\bibinfo{volume}{12}}, \bibinfo{pages}{024}
  (\bibinfo{year}{2006}{\natexlab{a}}), \eprint{hep-ph/0610206}.

\bibitem[{\citenamefont{Harlander
  et~al.}(2006{\natexlab{b}})\citenamefont{Harlander, Kant, Mihaila, and
  Steinhauser}}]{Harlander:2006rj}
\bibinfo{author}{\bibfnamefont{R.}~\bibnamefont{Harlander}},
  \bibinfo{author}{\bibfnamefont{P.}~\bibnamefont{Kant}},
  \bibinfo{author}{\bibfnamefont{L.}~\bibnamefont{Mihaila}}, \bibnamefont{and}
  \bibinfo{author}{\bibfnamefont{M.}~\bibnamefont{Steinhauser}},
  \bibinfo{journal}{JHEP} \textbf{\bibinfo{volume}{09}}, \bibinfo{pages}{053}
  (\bibinfo{year}{2006}{\natexlab{b}}), \eprint{hep-ph/0607240}.

\bibitem[{\citenamefont{Kunszt et~al.}(1994)\citenamefont{Kunszt, Signer, and
  Trocsanyi}}]{Kunszt:1993sd}
\bibinfo{author}{\bibfnamefont{Z.}~\bibnamefont{Kunszt}},
  \bibinfo{author}{\bibfnamefont{A.}~\bibnamefont{Signer}}, \bibnamefont{and}
  \bibinfo{author}{\bibfnamefont{Z.}~\bibnamefont{Trocsanyi}},
  \bibinfo{journal}{Nucl.Phys.} \textbf{\bibinfo{volume}{B411}},
  \bibinfo{pages}{397} (\bibinfo{year}{1994}), \eprint{hep-ph/9305239}.

\bibitem[{\citenamefont{Baikov and Chetyrkin}(2010)}]{Baikov:2010hf}
\bibinfo{author}{\bibfnamefont{P.}~\bibnamefont{Baikov}} \bibnamefont{and}
  \bibinfo{author}{\bibfnamefont{K.}~\bibnamefont{Chetyrkin}},
  \bibinfo{journal}{Nucl.Phys.} \textbf{\bibinfo{volume}{B837}},
  \bibinfo{pages}{186} (\bibinfo{year}{2010}), \bibinfo{note}{in memoriam
  Sergei Grigorievich Gorishny, 1958-1988}, \eprint{1004.1153}.

\bibitem[{\citenamefont{Siegel}(1980)}]{Siegel:1980qs}
\bibinfo{author}{\bibfnamefont{W.}~\bibnamefont{Siegel}},
  \bibinfo{journal}{Phys.Lett.} \textbf{\bibinfo{volume}{B94}},
  \bibinfo{pages}{37} (\bibinfo{year}{1980}).

\bibitem[{\citenamefont{Stockinger}(2005)}]{Stockinger:2005gx}
\bibinfo{author}{\bibfnamefont{D.}~\bibnamefont{St\"ockinger}},
  \bibinfo{journal}{JHEP} \textbf{\bibinfo{volume}{0503}}, \bibinfo{pages}{076}
  (\bibinfo{year}{2005}), \eprint{hep-ph/0503129}.

\bibitem[{\citenamefont{Signer and Stockinger}(2009)}]{Signer:2008va}
\bibinfo{author}{\bibfnamefont{A.}~\bibnamefont{Signer}} \bibnamefont{and}
  \bibinfo{author}{\bibfnamefont{D.}~\bibnamefont{St\"ockinger}},
  \bibinfo{journal}{Nucl. Phys.} \textbf{\bibinfo{volume}{B808}},
  \bibinfo{pages}{88} (\bibinfo{year}{2009}), \eprint{0807.4424}.

\bibitem[{\citenamefont{Tarasov et~al.}(1980)\citenamefont{Tarasov, Vladimirov,
  and Zharkov}}]{Tarasov:1980au}
\bibinfo{author}{\bibfnamefont{O.}~\bibnamefont{Tarasov}},
  \bibinfo{author}{\bibfnamefont{A.}~\bibnamefont{Vladimirov}},
  \bibnamefont{and} \bibinfo{author}{\bibfnamefont{A.}~\bibnamefont{Zharkov}},
  \bibinfo{journal}{Phys.Lett.} \textbf{\bibinfo{volume}{B93}},
  \bibinfo{pages}{429} (\bibinfo{year}{1980}).

\bibitem[{\citenamefont{Larin and Vermaseren}(1993)}]{Larin:1993tp}
\bibinfo{author}{\bibfnamefont{S.}~\bibnamefont{Larin}} \bibnamefont{and}
  \bibinfo{author}{\bibfnamefont{J.}~\bibnamefont{Vermaseren}},
  \bibinfo{journal}{Phys.Lett.} \textbf{\bibinfo{volume}{B303}},
  \bibinfo{pages}{334} (\bibinfo{year}{1993}), \eprint{hep-ph/9302208}.

\bibitem[{\citenamefont{Chetyrkin}(1997)}]{Chetyrkin:1997dh}
\bibinfo{author}{\bibfnamefont{K.~G.} \bibnamefont{Chetyrkin}},
  \bibinfo{journal}{Phys. Lett.} \textbf{\bibinfo{volume}{B404}},
  \bibinfo{pages}{161} (\bibinfo{year}{1997}),
  \eprint[http://arXiv.org/abs]{hep-ph/9703278}.

\bibitem[{\citenamefont{Vermaseren et~al.}(1997)\citenamefont{Vermaseren,
  Larin, and van Ritbergen}}]{Vermaseren:1997fq}
\bibinfo{author}{\bibfnamefont{J.~A.~M.} \bibnamefont{Vermaseren}},
  \bibinfo{author}{\bibfnamefont{S.~A.} \bibnamefont{Larin}}, \bibnamefont{and}
  \bibinfo{author}{\bibfnamefont{T.}~\bibnamefont{van Ritbergen}},
  \bibinfo{journal}{Phys. Lett.} \textbf{\bibinfo{volume}{B405}},
  \bibinfo{pages}{327} (\bibinfo{year}{1997}),
  \eprint[http://arXiv.org/abs]{hep-ph/9703284}.

\end{thebibliography}
\end{document}